\documentclass{article}
\usepackage{bbm}
\usepackage{graphicx}
\newtheorem{lemma}{Lemma}
\newtheorem{theorem}{Theorem}

\newtheorem{hypothesis}{Hypothesis}
\newtheorem{remark}{Remark}
\newtheorem{definition}{Definition}
\newcommand{\be}{\begin{eqnarray}}
\newcommand{\ee}{\end{eqnarray}}
\newcommand{\bee}{\begin{eqnarray*}}
\newcommand{\eee}{\end{eqnarray*}}
\newcommand{\R}{\mathbbm {R}}
\newcommand{\I}{\mbox {\sc 1}}
\newcommand{\asy}{{\mathcal O}}
\newcommand{\da}{d_A}

\begin{document}

\title {Nonlinear Schr\"odinger equations with multiple-well potential}

\author {Andrea SACCHETTI\\ 
Dipartimento di Scienze Fisiche, Informatiche e Matematiche\\ 
Universit\'a di Modena e Reggio Emilia, Modena, Italy\\
Via Campi 213/B, 41125 Modena - Italy\\ 
tel. +39.059.2055209, fax n. +39.059.2055584\\ 
email address: andrea.sacchetti@unimore.it\\
\\
and\\
\\ 
Centro S3, Istituto Nanoscienze, CNR\\ 
Via Campi 213/A, 41100 Modena, Italy}

\date {\today}

\maketitle

{\bf Abstract.} We consider the stationary solutions for a class of Schr\"odinger equations with a $N$-well potential and a nonlinear perturbation. \ By means of semiclassical techniques we prove that the dominant term of the ground state solutions is described by a $N$-dimensional Hamiltonian system, where the coupling term among the coordinates is a tridiagonal Toeplitz matrix. \ In particular we consider the case of $N=4$ wells, where we show the occurrence of spontaneous symmetry-breaking bifurcation effect. \ In particular, in the limit of large focusing nonlinearity we prove that the ground state stationary solutions consist of $N$ wavefunctions localized on a single well.

{\bf PACS number(s):} 05.45.-a, 02.30.Oz, 03.65.Sq, 03.75.Lm


\section {Introduction} \label {Sec1}

For a quantum system with ${\mathcal N}$ particles the Schr\"odinger equation is defined in a space with dimension $3{\mathcal N}+1$ and typically it is impossible to solve, neither analytically nor numerically even with today's supercomputers. \ However, assuming a mean field hypothesis, the $3{\mathcal N}+1$ dimensions linear system of  Schr\"odinger equation is approximated by a $3+1$ dimensions nonlinear Schr\"odinger equation. \ Although nonlinearity typically implies some new technical difficulties, the dimension is significantly reduced when compared with the original problem and this fact simplifies the study of  dynamics of quantum systems, independently of the total number ${\mathcal N}$ of particles. 

One of the most successful application of such an approach is the derivation of nonlinear Schr\"odinger equation for a Bose-Einstein condensate (BEC). \ Since its realization in diluted bosonic atomic gases \cite {A,BSH,HMEWC} the interest in studying the collective dynamics of macroscopic ensembles of atoms occupying the same quantum state is largely increased. \ The condensate typically consists of a few thousands to millions of atoms which are confined by a trapping potential and at temperature much smaller than some critical value, and a BEC is well described by the macroscopic wave function $\psi = \psi (x,t)$ whose time evolution is governed by a self-consistent mean field nonlinear Schr\"odinger equation  \cite {Pitaevskii}

Fos such a reason, in these last years there has been an increasing interest in the study of nonlinear Schr\"odinger equation with an external potential. \ In fact, many other interesting and current physical problems may be described by means of such a model, e.g. non-linear optics \cite {Joannopoulos}, semiconductors \cite {Mihalace}, and quantum chemistry \cite {Grecchi,Jona}, just to mention the most relevant. \ In particular, the mathematical research recently focused on the nonlinear Schr\"odinger equation (hereafter NLS) with double well potential. \ One of the most interesting feature of such a model is the spontaneous symmetry breaking phenomenon  \cite {JaWe,KKSW,S}, and recently a general rule in order to classify the kind of bifurcation has been obtained \cite {RS, Sacchetti} (see also \cite {Peli}). \ Much less is known for NLS with \emph {multiple well} potential. \ So far, few models with multiple wells have been considered, e.g. the model with three wells on a regular lattice \cite {Kapitula} (where a lattice means a sequence of points displaced along a straight line), and the model with four wells on the vertex of a regular square \cite {Wang}. \ The generic case with $N$ wells has not been yet studied. \ In fact, multiple-well potential represent the effect of \emph {small lattice} on, e.g., Bose-Einstein condensates; furthermore, they are also interesting in order to understand the transition to a lattice with infinitely many wells. 

In NLS problems with multi-well potentials the \emph {effective nonlinearity parameter} is usually given by the ratio between the strength of the nonlinear term and hopping matrix element between neighbour sites. \ The spontaneous symmetry breaking effect, and the associated localization phenomena, occurs when such a ratio is equal to a (finite) critical value. \ This fact has been seen, for instance, in the study of the localization effect in a gas of pyramidal molecules as the ammonia one \cite {Jona} or in the study of the Mott insulator-superfluid quantum phase transition \cite {Bl,Stoferle}. \ On the other side we also have to treat the problem of the validity of the $N$-mode approximation (where $N$ is the number of wells), obtained by restricting our analysis to the $N$-dimensional space associated to the first $N$ eigenvectors of the linear problem; in our approach we solve this problem considering the semiclassical limit of small $\hbar$. \ Since the hopping matrix element between neighbour sites is not fixed, but it is exponentially small when $\hbar$ goes to zero, then, in order to have a finite value for the effective nonlinearity parameter (if not then we simply have localization), we have to require that the strength of the nonlinear term should be exponentially small, too. \ Hence, in our model we introduce the multi-scale limit below in order to observe the bifurcation phenomena. \ We would point out that other multi-scale limits may be considered in order to obtain the validity of the $N$-mode approximation, e.g. one can consider the simultaneous limit of large distance between the wells and small nonlinear term as in \cite {Peli,KKSW}. \ The assumption of small $\hbar$ has the great advantage that, from a technical point of view, all the powerfull semiclassical results devoloped by Helffer and Sj\"ostrand in the 80' (see e.g. \cite {H}) are easily available when we consider the interaction between noighbour wells.

In this paper we consider a NLS with $N$ wells displaced on a regular lattice (even if the present analysis can be easily extended to the case on wells displaced on a regular grid, as discussed in an explicit example in Remark \ref {grid}). \ We'll show that such a problem can be reduced, up to a remainder term, to a finite dimensional system; to this end, instead of using some kind of Galerkin decomposition as in \cite {Kapitula}, we assume to be in the \emph {semiclassical limit}. \ In such a  way we can make use of some powerfull results of the semiclassical analysis \cite {H}. \ By means of such results and by making us of the Lyapunov-Schmidt reduction scheme we estimate the remainder term.

The finite-dimensional system we obtain is almost decoupled, in the sense that the coupling term which represents the interaction among the adjacent wells is associated to a tridiagonal Toeplitz matrix. \ Furthermore, it can be written in Hamiltonian form, where one of the coordinate is a cyclic coordinate.

We then consider in details the case of $N=4$ wells and we study the bifurcation picture when the strength of the nonlinear perturbation increases. \ As in the double well model we can see that the ground state stationary symmetric solution bifurcates giving arise to $4$ stationary solutions fully localized on a single well, and the kind of bifurcation satisfies the same rule as in the double well model. \ Actually, such a result may be generalized to any number $N$ of wells by means of a simple asymptotics argument. \ In particular, we focus our attention on the value of the \emph {effective nonlinearity parameter} at the bifurcation point in the case of $N=2,4,6$ and $8$ wells; indeed bifurcation phenomena is associated to the phase transition and we'll see that the results obtained by our model agree with the numerical experiment on the Bose-Hubbard model \cite {Stoferle}.

{\it Acknowledgments: I am very grateful to Reika Fukuizumi for useful discussions.}

\section {Description of the model}

Here, we consider the nonlinear Schr\"odinger (hereafter NLS) equations
\be
i \hbar \frac {\partial \psi }{\partial t} = H_0 \psi + \epsilon g(x) |\psi |^{2\sigma}  \psi , \ \psi (\cdot , t ) \in L^2 (\R^d , dx), \ \| \psi (\cdot ,t) \| =1 , \label {Eq1}
\ee
where $\epsilon \in \R$ and $\| \cdot \|$ denotes the $L^2 (\R^d ,dx)$ norm, 
\be
H_0 = - \frac {\hbar^2}{2m} \Delta + V , \ \  \ \Delta = \sum_{j=1}^d \frac {\partial^2}{\partial x_j^2} \, ,  \label {Eq2}
\ee
is the linear Hamiltonian with a \emph {multiple-well} potential $V(x)$, and $g(x) |\psi |^{2\sigma} $ is a nonlinear perturbation. \ For the sake of definiteness we assume the units such that $2m=1$. \ The semiclassical parameter $\hbar$ is such that $\hbar \ll 1$.

Here, we introduce the assumptions on the multiple-well potential $V$ and we collect some semiclassical results on the linear operator $H_0$.

\begin {hypothesis} \label {Hyp1}
Let $v(x) \in C^\infty_0 (\R^d ) $ be a smooth compact support function with a non degenerate minimum value at $x=0$:
\be
v (x) > v_{min} = v (0 ) , \ \ \forall x \in \R^d , \ x\not= 0  . \label {Eq3}
\ee

We consider multiple-well potentials of the kind
\be
V(x) = \sum_{j=1}^N v (x-x_j ) \label {Eq4}
\ee
for some $N>1$, where 
\bee
x_j = \left ( j \ell - \frac {N+1}{2} \ell , 0, \ldots , 0 \right ) \, 
\eee
and $\ell > 2 r$, where $r>0$ is such that 
\bee
{\mathcal C} \subseteq [-r,+r] \times \R^{d-1}  
\eee
and where ${\mathcal C}$ is the compact support of $v(x)$.

Hence, the multi-well potential $V(x)$ has exactly $N$ non degenerate minima at $x= x_j$, $j=1,2,\ldots , N$. 
\end {hypothesis}

\begin {remark}
It is a well known fact \cite {RS} that the Cauchy problem (\ref {Eq1}) is globally well-posedness for any $\epsilon \in \R$ small enough provided that 
\bee
\sigma < 
\left \{
\begin {array}{ll}
+\infty & \ \mbox { if } d<2 \\
\frac {1}{d-2} & \ \mbox { if } d >2
\end {array}
\right. \, . 
\eee
In such a case the conservation of the norm of $\psi (x,t)$ and of the energy 
\bee
{\mathcal E} (\psi ) = \langle \psi , H_0 \psi \rangle + \frac {\epsilon}{\sigma +1} \langle \psi^{\sigma +1} , g \psi^{\sigma +1} \rangle 
\eee
follows; furthermore we also have a priori estimate 
\be
\| \psi (\cdot , t )^{2\sigma} \| \le C \hbar^{-m} \label {Eq4Bis}
\ee
for some positive constants $C$ and $m$.
\end {remark}

\section {Analysis of the linear Schr\"odinger equation}

Now, making use of semiclassical analysis \cite {H} we look for the ground state of the linear Schr\"odinger equation 
\bee
H_0 \psi = \lambda \psi , \ \ \psi \in L^2 (\R^d ) \, . 
\eee
Let $\da (x,y)$ be the Agmon distance between two points $x$ and $y$, let 
\bee
S_0 = \inf_{i\not= j} \da (x_j , x_{i}) ;
\eee
then, by construction of the potential $V(x)$, it turns out that
\be
S_0 =  \da (x_j , x_{j+1}) , \ j=1, 2, \ldots , N -1\, , \label {Eq5} 
\ee
and 
\be 
S_0 < \da (x_j , x_{i}) \ \mbox { if } \ |i-j|>1 \, . \label {Eq5Bis}
\ee

Now, let $H_D$ be the Dirichlet realization of 
\be
H_D = - \hbar^2 \Delta + v \label {Eq6} 
\ee
on the ball $B_S (0)$ with center at $x=0$ and radius $S > 2 S_0$. \ Since the bottom of $v(x)$ is not degenerate, then the Dirichlet problem associated to the single-well trapping potential $v(x)$ has spectrum $\sigma (H_D)$ with ground state 
\bee
\lambda_D = v(0) + \sum_{j=1}^d \sqrt {\mu_j} \hbar + \asy (\hbar^2 ) \, , 
\eee
where $2\mu_j$ are the positive eigenvalues of the Hessian matrix $v'' (0)$, such that 
\bee
\mbox {dist} \left [ \lambda_D , \sigma (H_D) \setminus \{ \lambda_D \} \right ] > 2C \hbar 
\eee
for some $C>0$; the associated normalized eigenvector $\psi_D (x)$ is localized in a neighborhood of $x=0$ and it exponentially decreases as $\asy \left ( \hbar^{-m} e^{-\da (x)/\hbar} \right )$ for some $m>0$ and where $\da (x)$ is the Agmon distance between $x$ and the point $x=0$. 

The spectrum $\sigma (H_0)$ of $H_0$ contains exactly $N$ eigenvalues $\lambda_j$, $j=1,2,\ldots ,N$, such that 
\bee
\lambda_j - \lambda_D = \asy (e^{-\rho /\hbar }) 
\eee
for any $0< \rho < S_0$; this result is a conseguence of the fact that the multiple well potential $V(x)$ is given by a superposition of $N$ exactly equal wells displaced on a regular lattice. \ Furthermore  
\bee
\mbox {dist} \left [ \{ \lambda_j \}_{j=1}^N , \sigma (H_0) \setminus \{ \lambda_j \}_{j=1}^N \right ] > C \hbar .
\eee
Let $F$ be the eigenspace spanned by the eigenvectors $\psi_j$ associated to the eigenvalues $\lambda_j$. \ Then, the restriction $H_0|_F$ of $H_0$ to the subspace $F$ can be represented in the basis of orthonormalized vectors $\phi_j$ such that 
\be
\phi_j (x) - \varphi_j (x) = \asy \left ( e^{-\rho /\hbar } \right )\, ,\ j=1,2,\ldots , N, \ \varphi_j (x) = \psi_D (x-x_j), \label {Eq7Bis}
\ee
for any fixed $0< \rho < S_0 $; the eigenvector $\phi_j$ is localized in a neighborhood of the minima points $x_j$. \ More precisely, $H_0|_F$ in the basis $\phi_j (x)$, $j=1,2,\ldots , N$, is represented by the $N\times N$ matrix (see, e.g. Theorem 4.3.4 by \cite {H})
\be
\lambda_D \I_N + (w_{i,j})_{i,j=1}^N + \asy (\hbar^{\infty }) e^{-S_0/\hbar } \label {Eq7}
\ee
where
\be
w_{i,j} = 0 \, , \ \mbox { if } \ |i-j|\not= 1 , \label {Eq8}
\ee
and where 
\bee
w_{i,i+1} = - \beta 
\eee
is independent of $i=1, \ldots , N$ and it is such that (see Theorem 4.4.6 by \cite {H})
\be
\frac {1}{C} \hbar^{1/2} \le \beta e^{S_0/\hbar} \le C \hbar^{1-d/2} \,  . \label {Eq9}
\ee

\begin {remark}
Let $i = 1, \ldots , N-1$ be fixed and let $\Omega$ be an open set with smooth boundary such that $x_i \in \Omega$ and $x_j \notin \Omega$ for $i \not= j$, let 
\bee
E_i^{(a)} = \left \{ x \ : \ \da (x_i,x) + \da (x_{i+1,x}) \le S_0 +a \right \} 
\eee
for some $a>0$ sufficiently small. \ Let $\Gamma_i = \partial \Omega \cap E_i^{(a)}$, then 
\be
\beta = - \hbar^2 \int_{\Gamma_i} \left [ \psi_D (x-x_i) \frac {\partial  \psi_D (x-x_{i+1})}{\partial n} - \psi_D (x-x_{i+1}) \frac {\partial  \psi_D (x-x_i)}{\partial n} \right ] d S_\Gamma \, . \label {Eq12Bis}
\ee
The dominant term of $\beta$ is independent of $a$.

In particular, in dimension one, i.e. $d=1$, then it turns out that 
\bee
\beta = 2 \hbar^2 \psi_D \left ( \frac 12 \ell \right )  \psi_D' \left ( \frac 12 \ell \right ) \, . 
\eee
\end {remark}

Collecting all these results then we can conclude that

\begin {lemma} \label {Lemma1} 
Let $F$ be the eigenspace spanned by the eigenvectors $\psi_j$ associated to the eigenvalues $\lambda_j$ of $H_0$. \ Then, the restriction $H_0|_F$ of $H_0$ to the subspace $F$ can be represented in the basis of vectors $\phi_j$, localized on the $j-$th well and satisfying (\ref {Eq7Bis}), by the $N\times N$ tridiagonal Toeplitz matrix 
\be
T + \asy (\hbar^{\infty }) e^{-S_0/\hbar } \label {Eq10}
\ee
where
\be
T = - \beta {\mathcal T} + \lambda_D \I_N \label {Eq11}
\ee
and
\be
{\mathcal T} = \left (  \begin {array}{cccccccc}
0 & 1 & 0 & 0 & 0 & 0 & \cdots & 0 \\
1 & 0 & 1 & 0 & 0 & 0 & \cdots & 0  \\
0 & 1 & 0 & 1 & 0 & 0 & \cdots & 0 \\ 
0 & 0 & 1 & 0 & 1 & 0 & \cdots & 0 \\
\vdots & \vdots & \vdots & \ddots & \ddots & \ddots & \vdots & \vdots \\
0 & \cdots & 0 & 0 & 1 & 0 & 1 & 0  \\
0 & \cdots & 0 & 0 & 0 & 1 & 0 & 1  \\
0 & \cdots & 0 & 0 & 0 & 0 & 1 & 0 
\end {array}
\right )  \, . \label {Eq11BisBis}
\ee
where $\beta$ is the positive real number given by (\ref {Eq12Bis}), and satisfying (\ref {Eq9}) for some $C>0$.
\end {lemma}

From (\ref {Eq10}) it turns out that the eigenvalues $\lambda_j$ are given, up to a small correction $\asy (\hbar^\infty ) e^{-S_0/\hbar }$, by the eigenvalues $\mu_j$ of the matrix $T$. \ To this end we recall that the $N$ eigenvalues of the tridiagonal Toeplitz matrix $T$ are given by \cite {M} 
\bee
\mu_j = \lambda_D - 2 \beta \cos \left ( \frac {j \pi}{N+1} \right ) \, , \ j=1,2,\ldots , N , 
\eee
with associated eigenvectors 
\bee
(v_j)_k = \sin \left ( \frac { k j \pi}{N+1} \right ) . 
\eee
In order to normalize the eigenvector we remark that
\bee
{\sum_{k=1}^N \sin^2 \left ( \frac { k j \pi}{N+1} \right )} = \frac {N+1}{2}
\eee
From this fact and from Lemma \ref {Lemma1} we then conclude that

\begin {lemma} \label {Lemma2}
The first $N$ eigenvalues $\lambda_j$ of $H_0$ are given by 
\bee
\lambda_j  = \lambda_D - 2 \beta \cos \left ( \frac {j \pi}{N+1} \right ) + \asy (\hbar^{\infty }) e^{-S_0/\hbar }\, , \ j=1,2,\ldots , N , 
\eee
and the associated normalized eigenvector are given by
\be
\psi_j (x) = \sum_{k=1}^N \alpha_{j,k} \varphi_k (x)  + \asy (\hbar^{\infty }) e^{-S_0/\hbar }\, , \ j=1,2,\ldots , N , \label {Eq11Bis}
\ee
where
\bee
\alpha_{j,k}= \alpha_{k,j}= \sqrt {\frac {2}{N+1}}  \sin \left ( \frac { k j \pi}{N+1} \right ) \ \ \mbox { and } \ \ \varphi_k (x) = \psi_D (x-x_k).
\eee
\end {lemma}

In the Appendix we'll consider in detail the case of $N=2,3,4$ wells.

\begin {remark} \label {grid}
We can immediately extend such an analysis to multiple-well potentials of the form $V(x) = \sum_j v(x-x_j)$ where $x_j \in \R^d$ are points on a regular grid. \ We study, for argument sake's, the model in $\R^2$ considered by \cite {Wang} where the potential $V(x)$ has $4$ wells with minima on the points 
\bee
x_1 =(1,1);\ x_2=(1,-1);\ x_3 =(-1,1);\ x_4 = (-1,-1) \, . 
\eee
In such a case we have that (recalling that $w_{i,j}=w_{j,i}$)
\bee
w_{1,2} = w_{1,3} = w_{2,4} = w_{4,3} =-\beta \ \mbox { and } \ w_{i,i}=0,\ w_{1,4}=w_{2,3} =0 \, 
\eee
and the matrix $T$ takes the form
\bee
T = 
\left ( 
\begin {array}{cccc}
\lambda_D & - \beta & - \beta & 0 \\ 
- \beta & \lambda_D & 0 & - \beta \\ 
- \beta & 0 & \lambda_D & - \beta \\
0 & - \beta & - \beta & \lambda_D 
\end {array}
\right ) 
\eee
with eigenvalues (and associated normalized eigenvectors $(v_j)_k$)
\bee
\mu_1 &=& \lambda_D - 2 \beta \, , \ v_1 =\left ( \frac 12 ,   \frac 12 ,  \frac 12 , \frac 12 \right ) \\ 
\mu_2 &=& \lambda_D  \, , \ v_2 =\left ( -\frac {1}{\sqrt 2} ,  0 , 0 , \frac {1}{\sqrt 2} \right ) \\ 
\mu_3 &=& \lambda_D  \, , \ v_3 =\left ( 0, -\frac {1}{\sqrt 2} ,  \frac {1}{\sqrt 2},0 \right ) \\
\mu_1 &=& \lambda_D - 2 \beta \, , \ v_1 =\left ( \frac 12 ,   -\frac 12 , - \frac 12 , \frac 12 \right )
\eee
in agreement with \cite {Wang}.
\end {remark}

\section {The $N$-mode approximation for the NLS equation}

Let $\psi$ be the normalized solution of the NLS equation (\ref {Eq1}) written in the form
\be
\psi (x,t) = \sum_{j=1}^N c_j (t) \psi_j (x) + \psi_c (x,t) \, . \label {Eq12}
\ee
By substituting (\ref {Eq12}) into (\ref {Eq1}) and, projecting on the eigenspaces spanned by the eigenvectors $\psi_j$ and on the orthogonal eigenspace, we obtain the following system of differential equations:
\be
\left \{
\begin {array}{lcl}
i \hbar \dot c_j &=& \lambda_j c_j  + \epsilon \langle \psi_j (\cdot ) , g(\cdot ) |\psi (\cdot ,t)|^{2\sigma} \psi (\cdot ,t) \rangle \\ 
i \hbar \dot \psi_c &=& H_0 \psi_c + \epsilon \Pi_c g(x) |\psi (x,t)|^{2\sigma} \psi (x,t)
\end {array}
\right. \label {Eq13}
\ee
where we set 
\bee
\Pi = \sum_{j=1}^N \langle \psi_j, \cdot \rangle \psi_j \ \ \mbox { and } \ \ \Pi_c = \I - \Pi \, .
\eee
By substituting (\ref {Eq11Bis}) in (\ref {Eq13}) we obtain 
\bee
i \hbar \dot c_j 
&=& \lambda_j c_j + \epsilon \sum_k \bar \alpha^{j,k} \langle \varphi_k (\cdot ), g(\cdot ) |\psi (\cdot , t)|^{2\sigma} \psi (\cdot , t) \rangle + r_k \\ 
&=& \lambda_j c_j + \epsilon \sum_{k} \bar \alpha^{j,k} |d_k|^{2\sigma } d_k \langle \varphi_k (\cdot ), g(\cdot ) |\varphi_k (\cdot )|^{2\sigma} \varphi_k (\cdot ) \rangle + r_k \\ 
&=& (\lambda_D + \omega_j) c_j + \epsilon \sum_{k}  \alpha^{j,k} C_k |d_k|^{2\sigma } d_k  +  r_k
\eee
where we set 
\be
d_k := d_k (t) = \sum_{j=1}^N c_j (t) \alpha_{j,k} \, , \label {Eq14Bis}
\ee
\bee
r_k = \epsilon \langle \varphi_k (\cdot ), g(\cdot ) |\psi_c (\cdot ,t )|^{2\sigma } \psi_c (\cdot , t) \rangle + \epsilon \asy (\hbar^\infty ) e^{-S_0/\hbar} \, , 
\eee
\bee
\omega_j = - 2 \beta \cos \left ( \frac {j \pi}{N+1} \right ) \ \mbox { and } \ C_k = \langle \varphi_k^{\sigma +1} (\cdot ), g(\cdot ) \varphi_k^{\sigma +1} (\cdot )  \rangle \, , 
\eee
since $\langle \varphi_k , |\varphi_m |^{2\sigma } \varphi_m \rangle = \asy (\hbar^\infty ) e^{-S_0/\hbar}$ for $m \not= k$ and $ \alpha_{j,k} = \bar  \alpha_{j,k} = \alpha_{k,j}$, and where we make use of the \emph {a priori} estimate (\ref {Eq4Bis}) of the norm of $|\psi |^{2\sigma}$. \ We should underline that, by construction and by Lemma \ref {Lemma2}, it follows that
\bee
\psi (x,t) = \sum_{k=1}^N d_k (t) \varphi_k (x) + \psi_c (x,t) + \asy (\hbar^\infty ) e^{-S_0/\hbar } \, . 
\eee

If we denote by
\bee
A=(\alpha_{j,k}), \ C=(C_k),\ d=(d_k), \ \lambda = \mbox {diag} (\lambda_j), \ r = (r_k )_k \, , 
\eee
then the above equation takes the form (with abuse of notation)
\bee
i \hbar A^{-1} \dot d  = \Lambda A^{-1} d + \epsilon  C A^{-1} |d|^{2\sigma} d + r
\eee
that is
\be
i \hbar \dot d_k = ( T d )_k + \epsilon \tilde C_k |d_k|^{2\sigma} d_k + \tilde r  \label {Eq15}
\ee
since
\bee
T = A\Lambda A^{-1}
\eee
and where we set 
\bee
\tilde C = A C A^{-1} \ \mbox { and } \ \tilde r = A r . 
\eee

\begin {definition}
We call {\bf $N$-mode approximation} for the NLS equation the system of ODEs obtained by neglecting the remainder term $\tilde r$
\be
i \hbar \dot d_k = ( T d )_k + \epsilon \tilde C_k |d_k|^{2\sigma} d_k \, , \ k=1,2,\ldots , N, \label {Eq15Bis}
\ee
where $\tilde C_k $ are real-valued constant, and with the normalization condition
\be
\sum_{k=1}^N |d_k (t) |^2 = 1\, . \label {Eq16}
\ee
\end {definition}

The validity of the $N$-mode approximation for large times is, in general, an open problem . \ So far it has been proved \cite {BS} that if the state is initially prepared on the space spanned by the $N$ linear eigenvectors then remainder term $\psi_c (\cdot , t)$ is norm bounded by an exponentially small term for times of order $\beta^{-1}$, furthermore the difference $c_j(t)-d_j(t)$, between the coefficients of the solution of the NLS equation and the solutions of the $N$-mode approximation, has the same exponentially small estimate for times of order $\beta^{-1}$ too. \ This result can be extented for larger times of the order $e^{\beta^{-1}}$ under further technical assumptions. \ Non linear systems (\ref {Eq15Bis}) can be studied by means of dynamical systems methods, see \cite {GMS} for the $N=2$ wells model. 

Concerning the study of the stationary solutions $\psi (x,t) = e^{i\omega t} \psi (x)$ has been proved by \cite {RS} that the $2$-mode approximation gives the stationary solutions for the NLS, up to an exponentially small error, furthermore the orbital stability of the stationary solutions is proved; the same argument may apply to the $N$-mode approximation for any $N \ge 2$ proving that the stationary solutions of equations (\ref {Eq15Bis}) and (\ref {Eq16}) give, up to an exponentially small error $\asy (e^{-\rho /\hbar })$, for any $0<\rho <S_0$, the stationary solution of the NLS (\ref {Eq1}). \ However, we don't dwell here on these details.

For instance, in the case of two symmetric wells, i.e., $N=2$ then (\ref {Eq15Bis}) takes the form (in agreement with \cite {Sacchetti})
\bee
\left \{
\begin {array}{lcl}
i \hbar \dot d_1 &=&  \lambda_D d_1 - \beta d_2 + \epsilon \tilde C_1 |d_1|^2 d_1  \\ 
i \hbar \dot d_2 &=&  \lambda_D d_2 - \beta d_1 + \epsilon \tilde C_2 |d_2|^2 d_2  
\end {array}
\right.
\eee
In the case of three symmetric wells, i.e., $N=3$ then (\ref {Eq15Bis}) takes the form (in agreement with \cite {Kapitula})
\bee
\left \{
\begin {array}{lcl}
i \hbar \dot d_1 &=&  \lambda_D d_1 - \beta d_2 + \epsilon \tilde C_1 |d_1|^2 d_1  \\ 
i \hbar \dot d_2 &=&  \lambda_D d_2 - \beta d_1 - \beta d_3 + \epsilon \tilde C_2 |d_2|^2 d_2\\ 
i \hbar \dot d_3 &=&  \lambda_D d_3 - \beta d_2 + \epsilon \tilde C_3 |d_3|^2 d_3 
\end {array}
\right.
\eee
Finally, in the case of four symmetric wells, i.e., $N=4$ then (\ref {Eq15Bis}) takes the form
\be
\left \{
\begin {array}{lcl}
i \hbar \dot d_1 &=&  \lambda_D d_1 - \beta d_2 + \epsilon \tilde C_1 |d_1|^2 d_1  \\ 
i \hbar \dot d_2 &=&  \lambda_D d_2 - \beta d_1 - \beta d_3 + \epsilon \tilde C_2 |d_2|^2 d_2  \\ 
i \hbar \dot d_3 &=&  \lambda_D d_3 - \beta d_2 - \beta d_4 + \epsilon \tilde C_3 |d_3|^2 d_3  \\ 
i \hbar \dot d_4 &=&  \lambda_D d_4 - \beta d_3  + \epsilon \tilde C_4 |d_4|^2 d_4 
\end {array}
\right. \label {pippo}
\ee

\subsection {Hamiltonian form of the $N$-mode approximation}
If we set
\bee
d_k = \sqrt {q_k} e^{i \theta_k}, \ \ q_k \in [0,1]\, , \ \theta_k \in [0, 2 \pi )\, ,  
\eee
then, by means of a straightforward calculation, it follows that (\ref {Eq15Bis}) takes the Hamiltonian form
\be
\left \{ 
\begin {array}{lcl}
\hbar \dot q_k &=& \frac {\partial {\mathcal H}}{\partial \theta_k} = - 2 \beta \sum_{j=1}^N {\mathcal T}_{j,k} \sqrt {q_k q_j} \sin (\theta_j - \theta_k) \\ 
\hbar \dot \theta_k &=& - \frac {\partial {\mathcal H}}{\partial q_k} = - \left [ \lambda_D - \beta \sum_{j=1}^N {\mathcal T}_{j,k} \sqrt {\frac {q_j}{q_k}} \cos (\theta_j - \theta_k) + \epsilon \tilde C_k q_k^\sigma \right ] 
\end {array}
\right. \,  \label {Eq17}
\ee
with Hamiltonian function
\be
{\mathcal H} &:=& {\mathcal H} (q_1, \ldots , q_N , \theta_1 , \ldots , \theta_N) \nonumber \\ 
&=& \lambda_D \sum_{k=1}^N q_k - \beta \sum_{j,k=1}^N {\mathcal T}_{k,j} \cos (\theta_j-\theta_k) \sqrt {q_j q_k} + \epsilon \frac {1 }{\sigma+1} \sum_{k=1}^N \tilde C_k q_k^{\sigma+1} \nonumber \\ 
&=& \lambda_D \sum_{k=1}^N q_k - 2 \beta \sum_{k=1}^{N-1} \cos (\theta_{k+1}-\theta_k) \sqrt {q_{k+1} q_k} + \epsilon \frac {1 }{\sigma+1} \sum_{k=1}^N \tilde C_k q_k^{\sigma+1} \, . \label {Eq18}
\ee
The normalization condition (\ref {Eq16}) takes the form
\be
\sum_{k=1}^N q_k =1 \, , \label {Eq16Bis}
\ee
furthermore the Hamiltonian function ${\mathcal H}$ is an integral of motion; i.e. 
\be
{\mathcal H} \left [ q_1 (t), \ldots , q_N (t), \theta_1 (t), \ldots , \theta_N(t) \right ] = const. \, . \label {Eq18Bis}
\ee

\subsection {Reduced Hamiltonian} 
We make use of the fact that $\sum_{k} q_k=1$ in order to reduce from $N$ to $N-1$ the degree of freedom of the Hamiltonian system (\ref {Eq18}). \ We consider the canonical transformation $(q,\theta ) \to (Q , \Theta )$ defined as 
\bee
Q_h = \sum_{k=1}^h q_k \, 
\eee
where $Q_1 \in [0,1]$ and $Q_h \in [0, 1-Q_{h-1}]$, $h=2, \ldots , N-1$. \ The inverse transformation is defined as
\bee
q_1 = Q_1 \ \mbox { and } \ q_h = Q_h - Q_{h-1} \, , \ h=2,\ldots , N\, .
\eee
The associated transformation on the conjugate variable $\theta$ is then given by 
\bee
\Theta_N = \theta_N \ \mbox { and } \ \Theta_h = \theta_h - \theta_{h+1}  \, , \ h=2,\ldots , N\, , 
\eee
with inverse 
\bee
\theta_h = \sum_{k=h}^N \Theta_k \, . 
\eee
In the coordinates $(Q, \Theta )$ the Hamiltonian system takes the form
\be
\left \{ 
\begin {array}{lcl}
\hbar \dot Q_k &=& \frac {\partial {\mathcal K}}{\partial \Theta_k} \\ 
\hbar \dot \Theta_k &=& - \frac {\partial {\mathcal K}}{\partial Q_k}
\end {array}  
\right.
 \, , \ k=1,\ldots , N\, ,   \label {EqCan1}
\ee
where the new Hamiltonian denoted by ${\mathcal K}$ is given by
\bee
{\mathcal K} &=& {\lambda_D} Q_N  -2 \beta \cos (\Theta_1) \sqrt {(Q_2-Q_1) Q_1} - 2 \beta \sum_{k=2}^{N-1} \cos (\Theta_k) \sqrt {(Q_{k+1}-Q_k)(Q_k-Q_{k-1})} + \\ 
&& \ \ + {\epsilon} \frac {1}{\sigma +1} \left [ \tilde C_1 Q_1^{\sigma+1} + \sum_{k=2}^N \tilde C_k (Q_k - Q_{k-1})^{\sigma +1} \right ] 
\eee
It turns out hat $\Theta_N$ is a cyclic coordinate (indeed $Q_N = const.=1$), then the Hamiltonian system if finally given by
\be
\left \{ 
\begin {array}{lcl}
\hbar \dot Q_k &=& \frac {\partial {\mathcal K}^\star}{\partial \Theta_k} \\ 
\hbar \dot \Theta_k &=& - \frac {\partial {\mathcal K}^\star}{\partial Q_k}
\end {array} \right. \, , \ k=1,\ldots , N-1\, ,   \label {EqCan2}
\ee
with Hamiltonian function 
\bee
{\mathcal K}^\star &=& {\lambda_D}  -2 \beta \cos (\Theta_1) \sqrt {(Q_2-Q_1) Q_1} - 2 \sum_{k=2}^{N-2} \cos (\Theta_k) \sqrt {(Q_{k+1}-Q_k)(Q_k-Q_{k-1})} - \\ 
&& \ \ - 2 \beta \cos (\Theta_{N-1}) \sqrt {(1-Q_{N-1})(Q_{N-1 }-Q_{N-2})}+ \\ 
&& \ \ + {\epsilon} \frac {1}{\sigma +1} \left [ \tilde C_1 Q_1^{\sigma+1} + \sum_{k=2}^{N-1} \tilde C_k (Q_k - Q_{k-1})^{\sigma +1} + \tilde C_N (1 - Q_{N-1})^{\sigma +1}\right ] \, . 
\eee

\section {Stationary solutions}

Now, we look for the normalized stationary solutions of the form $\psi (x,t) = e^{i \omega t} \psi (x)$. \ In terms of 
$N$-mode approximation (\ref {Eq18}) it consists of looking for the solution of the system of equations 
\be
\left \{ 
\begin {array}{lcl}
0 &=& \frac {\partial {\mathcal H}}{\partial \theta_k} \\ 
\hbar \omega &=& - \frac {\partial {\mathcal H}}{\partial q_k}
\end {array}
\right. \, , \label {Eq20}
\ee
That is, equation (\ref {Eq20}) and the normalization condition lead us to the following system
\be
\left \{
\begin {array}{lcl}
- 2\sin (\theta_{k+1}- \theta_k )\sqrt {q_{k+1} q_k} + 2 \sin (\theta_{k} - \theta_{k-1} ) \sqrt {q_{k-1} q_k} &=& 0  \\
- 2 \sin (\theta_{N} - \theta_{N-1} ) \sqrt {q_{N} q_{N-1}} &=& 0 \\ 
+ 2 \sin (\theta_{2} - \theta_1 ) \sqrt {q_{2} q_1} &=& 0 \\ 
{\lambda_D} - \beta \left [ \cos (\theta_{k+1} - \theta_k ) \sqrt {\frac {q_{k+1}}{q_k}} + \cos (\theta_{k} - \theta_{k-1} ) \sqrt {\frac {q_{k-1}}{q_k}} \right ] + {\epsilon} \tilde C_k q_k^\sigma &=& - \hbar \omega \\
{\lambda_D} - \beta \cos (\theta_{N} - \theta_{N-1} ) \sqrt {\frac {q_{N-1}}{q_N}}  + {\epsilon} \tilde C_N q_N^\sigma &=& - \hbar \omega  \\ 
{\lambda_D} - \beta \cos (\theta_{2} - \theta_{1} ) \sqrt {\frac {q_{2}}{q_1}}  + {\epsilon} \tilde C_1 q_1^\sigma &=& - \hbar \omega  \\ 
q_1 + q_2 + \ldots + q_N &=& 1  
\end {array}
\right.  \label {Eq22}
\ee
where $k=2, \ldots , N-2$. 

\begin {remark} \label {Nota1}
We may remark that from the first three equations of ( \ref {Eq22}) it follows that 
\bee
\sin (\theta_{k+1} - \theta_k ) \sqrt {q_{k+1}q_k } = 0, \ \mbox { for any } k=1,\ldots , N-1 
\eee
Furthermore, from (\ref {Eq15Bis}) the stationary solutions where $d_1 \equiv 0$ and $d_N \equiv 0$ are not admitted, hence $|d_k|<1$ for any $k$. \ Then $q_1,q_N \in (0,1)$ and $q_k \in [0,1)$ for any $k=2,\ldots, N-1$. \ For what concern the phases $\theta_k \in [0,2\pi )$ they are defined up to a gauge term, then we can always assume that, e.g., $\theta_1 =0$.
\end {remark}

\subsection {Four wells} The stationary problem in the case of two wells and three wells have been already studied \cite {RS, Kapitula, Sacchetti}. \ We restrict our analysis to the four-well case. \ For the sake of definetness we consider the four-well ($N=4$) model where we assume that $g(x) $ is a constant function, hence $\tilde C_k $ is independent of $k$ and thus we set $C:=\tilde C_k = C_k$. \ In such a case (\ref {Eq22}) takes the form 
\be
\left \{
\begin {array}{lcl}
\sin (\theta_{2} - \theta_1 ) \sqrt {q_{2} q_1} &=& 0  \\
\sin (\theta_{3} - \theta_2 ) \sqrt {q_{3}q_2 } &=& 0  \\
\sin (\theta_{4} - \theta_3 ) \sqrt {q_{4}q_3 } &=& 0  \\
\frac {\lambda_D}{\beta} - \cos (\theta_{2} - \theta_{1} ) \sqrt {\frac {q_{2}}{q_1}}+ \eta q_1^\sigma &=& - \omega \hbar /\beta  \\ 
\frac {\lambda_D}{\beta} - \frac {1}{\sqrt {q_2}} \left [ \cos (\theta_{3} - \theta_2 ) \sqrt {q_{3}} + \cos (\theta_{2} - \theta_{1} ) \sqrt {q_{1}} \right ] + \eta q_2^\sigma &=& - \omega \hbar /\beta  \\
\frac {\lambda_D}{\beta} - \frac {1}{\sqrt {q_3}} \left [ \cos (\theta_{4} - \theta_3 ) \sqrt {q_{4}} + \cos (\theta_{3} - \theta_{2} ) \sqrt {q_{2}} \right ] + \eta q_3^\sigma &=& - \omega \hbar /\beta  \\
\frac {\lambda_D}{\beta} - \cos (\theta_{4} - \theta_{3} ) \sqrt {\frac {q_{3}}{q_4}} + \eta q_4^\sigma &=& - \omega \hbar /\beta  \\ 
q_1 + q_2 + q_3 +q_4 &=& 1  
\end {array}
\right.  \label {Eq24FourBis}
\ee
where we set 
\bee
\eta = \frac {\epsilon}{\beta} C \, .
\eee

First of all we underline that $q_2=0$ (and similarly $q_3=0$) cannot be a solution of such a system; indeed if $d_2 \equiv 0$ is a stationary solution of the $4$-mode approximation (\ref {pippo}), then (\ref {pippo}) reduces to
\bee
\left \{
\begin {array}{lcl}
i \hbar \dot d_1 &=&  \lambda_d d_1 + \epsilon C |d_1|^2 d_1  \\ 
0 &=&   - \beta d_1 - \beta d_3   \\ 
i \hbar \dot d_3 &=&  \lambda_d d_3 - \beta d_4 + \epsilon C |d_3|^2 d_3  \\ 
i \hbar \dot d_4 &=&  \lambda_d d_4 - \beta d_3  + \epsilon C |d_4|^2 d_4 
\end {array}
\right.
\eee
form which follows $d_1=d_3$ and then $d_4=0$, which is not possible since Remark \ref {Nota1}.

\begin {remark}
We can extend such an argument to any $N$: if $N$ is an even integer and positive number then the stationary solution are given for $q_k \in (0,1)$, for any $k$. \ If $N$ is an odd integer and positive number then the stationary solutions are given for $q_1,q_N \in (0,1)$ and $q_k \in [0,1)$, $k=2, \ldots , N-1$, that is $q_k =0$, for $k \not= 1,N$, may be admitted values for stationary solutions.
\end {remark}

Since the stationary solutions are given for $q_k \in (0,1)$ then it follows that
\bee
\theta_{k+1}-\theta_k = 0, \pi 
\eee
and we obtain a family of $8$ systems of equations
\be
\left \{
\begin {array}{lcl}
- (-1)^j \sqrt {\frac {q_{2}}{q_1}}+ \eta q_1^\sigma &=& \Omega  \\ 
- \frac {1}{\sqrt {q_2}} \left [ (-1)^\ell \sqrt {q_{3}} + (-1)^j \sqrt {q_{1}} \right ] + \eta q_2^\sigma &=& \Omega  \\
- \frac {1}{\sqrt {q_3}} \left [ (-1)^m \sqrt {q_{4}} + (-1)^\ell \sqrt {q_{2}} \right ] + \eta q_3^\sigma &=& \Omega  \\
- (-1)^m \sqrt {\frac {q_{3}}{q_4}} + \eta q_4^\sigma &=& \Omega \\ 
q_1 + q_2 + q_3 +q_4 &=& 1  
\end {array}
\right.  \, , \ j,m,\ell =1,2. \label {Eq24FourTer}
\ee
where  we set
\bee
\Omega = -\frac {\lambda_D+ \hbar \omega}{\beta}  \, . 
\eee

\subsection {Symmetric and antisymmetrical solutions} 

\begin{figure} [ht]
\begin{center}
\includegraphics[height=10cm,width=10cm]{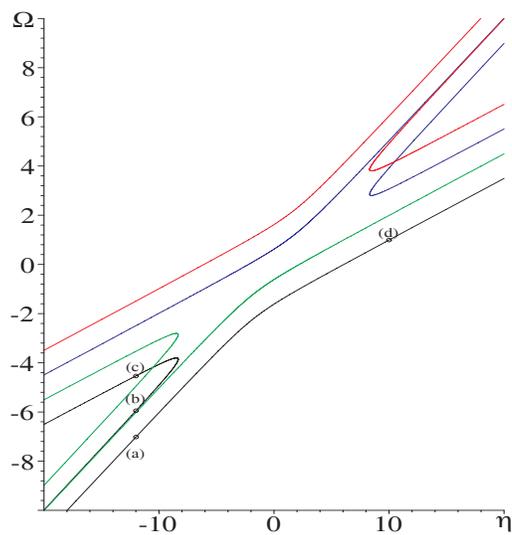}
\end{center}
\caption{\label {Fig4} Here we plot the energy $\Omega$ corresponding to symmetrical/antisymmetrical solutions versus $\eta$ in the case of cubic nonlinearity (i.e. $\sigma =1$). \ Black line corresponds to the choice $j=\ell =2$, and blue line corresponds to the choice $j=1$ and $\ell=2$, that is both graphs correspond to \emph {symmetric} solutions; green line corresponds to the choice $j=2$ and $\ell=1$, ans, finally, red line corresponds to the choice $j=\ell =1$, that is both graphs correspond to \emph {antisymmetrical} solutions.}
\end{figure}

In order to find symmetric and antisymmetrical solutions we set
\bee
q = q_1 = q_4 \ \mbox { and } \ p= q_2 =q_3 
\eee
In such a case it follows that $m=j$, that is (\ref {Eq24FourTer}) reduces to 
\be
\left \{
\begin {array}{lcl}
- (-1)^j \sqrt {\frac {p}{q}}+ \eta q^\sigma &=& \Omega  \\ 
- \frac {1}{\sqrt {p}} \left [ (-1)^\ell \sqrt {p} + (-1)^j \sqrt {q} \right ] + \eta p^\sigma &=& \Omega  \\
2q + 2p &=& 1  
\end {array}
\right.  \, , \ j,\ell =1,2. \label {Eq24FourQuaTer}
\ee
where to $j=2$ corresponds \emph {symmetric solutions}, while to $j=1$ corresponds \emph {antisymmetrical solutions}.

We see that the exchange $j \to 3-j$ reduces to the same system provided that $\eta \to - \eta $, $E \to - E$ and $\ell \to 3- \ell$; thus we can choose, for argument's sake, $j=2$ obtaining
\bee
\left \{
\begin {array}{lcl}
-  \sqrt {\frac {p}{q}}+ \eta q^\sigma &=& E  - \frac 12 (-1)^\ell\\ 
 -  \sqrt {\frac {q}{p}}  + \eta p^\sigma &=& E + \frac 12 (-1)^\ell  \\
2q + 2p &=& 1  
\end {array}
\right.  \, , \ \ell =1,2. 
\eee
where we set 
\bee
\Omega =E - \frac 12 (-1)^\ell \, .
\eee

It immediately follows that the exchange $\ell \to 3-\ell$ reduces to the same system provided we perform the exchange $p \leftrightarrow q$ and $E \to E+ \frac 12 (-1)^\ell$. \ Therefore we can choose, for argument's sake, $\ell =2$ and the system takes the final form
\be
\left \{
\begin {array}{lcl}
-  \sqrt {\frac {\frac {1}{2} - q }{q}}+ \eta q^\sigma &=& E  - \frac 12 \\ 
 -  \sqrt {\frac {q}{{\frac {1}{2} - q }}}  + \eta \left ( {\frac {1}{2} - q }\right )^\sigma &=& E + \frac 12 
\end {array}
\right.  \, . \label {Eq25}
\ee

By means of a straightforward calculation we can obtain $\eta$ and $\Omega$ as functions on $q$ and plot $\Omega$ versus $\eta$. 

\begin{figure}[ht]
\begin{center}
\includegraphics[height=8cm,width=8cm]{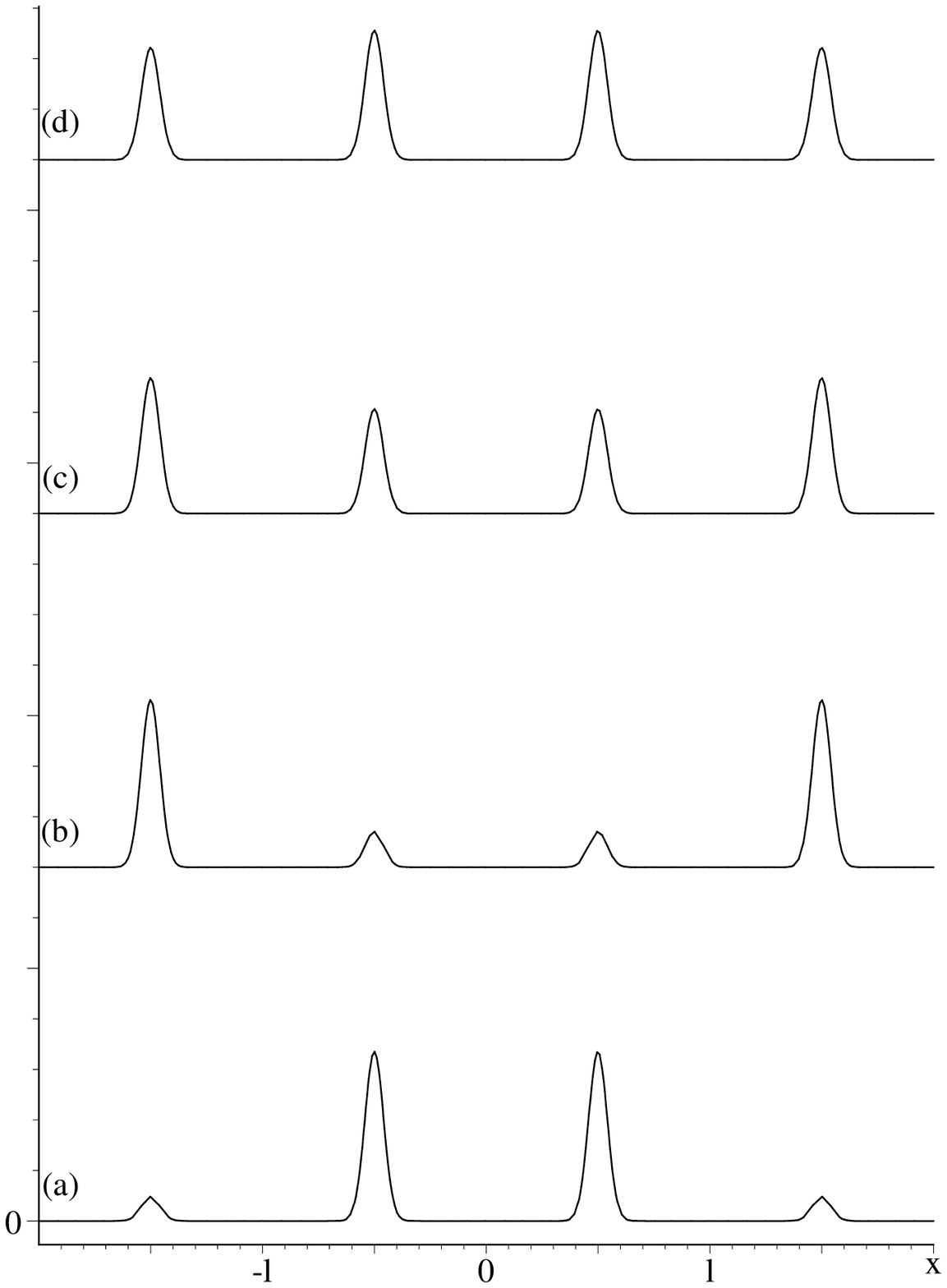}
\end{center}
\caption{\label {Fig5} Here we plot the absolute value of the symmetric wavefunctions corresponding to the case $j=\ell =2$ for $\eta =-12$ and $\eta =10$, in the case of cubic nonlinearity. \ For $\eta = -12 < - \eta^1$ we have a wavefunction (a), which is the continuation of the unperturbed one, plus two new wavefunctions (b) and (c) associated to branches coming from a  saddle point; for $\eta =10$ we only have the wavefunction (d),  which is the continuation of the unperturbed one.}
\end{figure}

In Figure \ref {Fig4} we consider the value of $\Omega$, as function of $\eta$, corresponding to symmetric/antisymmetrical stationary solutions in the cubic case, where $\sigma=1$. \ We may see that there exists a critical value  $ \eta^1 = 8.324 $ such that for any $|\eta | < \eta^1$ we only have $4$ symmetric/antisymmetrical stationary solutions as in the linear case (where $\eta =0$). \ At $\eta = \pm \eta^1$ saddle points occur and new branches of symmetric/antisymmetrical stationary solutions arise. \ The points denoted by (a), (b), and (c) correspond to values of $\Omega$ for $\eta =-12 < - \eta^1$ and where $j=\ell =2$; the point denoted by (d) corresponds to the unique value of $\Omega$ for $\eta =10$ and where $j=\ell =2$. \ In particular we have that (see also Figure \ref {Fig5})
\be
\begin {array}{llll}
(a) & \eta = -12 & \Omega = -7.021 & q = 0.010  \\ 
(b) & \eta =- 12 & \Omega = -5.952 & q = 0.478  \\ 
(c) & \eta = -12 & \Omega = -4.534 & q = 0.314  \\ 
(d) & \eta = 10  & \Omega = 0.990  & q = 0.214 
\end {array}
\label {Eqpippo}
\ee
We may remark that in the limit $|\eta | \to + \infty$, that is for large nonlinearity, then the wavefunctions (a) and (b) are, respectively, fully localized on the internal and external two wells, while the wavefunctions (c) and (d) are equally distributed on the four wells.

\begin {remark}
We should remark that the same picture occurs even for other values of $\sigma$; for instance in the case of $\sigma =2$ the saddle point occurs at $\eta =\eta^2 = -16.648$, in the case of $\sigma =3$ the the saddle point occurs at $\eta =\eta^3 = -38.775$. \ In general, computing $\eta^\sigma$ for higher values of $\sigma$ the following rule appears: $\eta^{\sigma+1} / \eta^\sigma \sim 2$ for large $\sigma$.
\end {remark}

\begin{figure}[ht]
\begin{center}
\includegraphics[height=10cm,width=10cm]{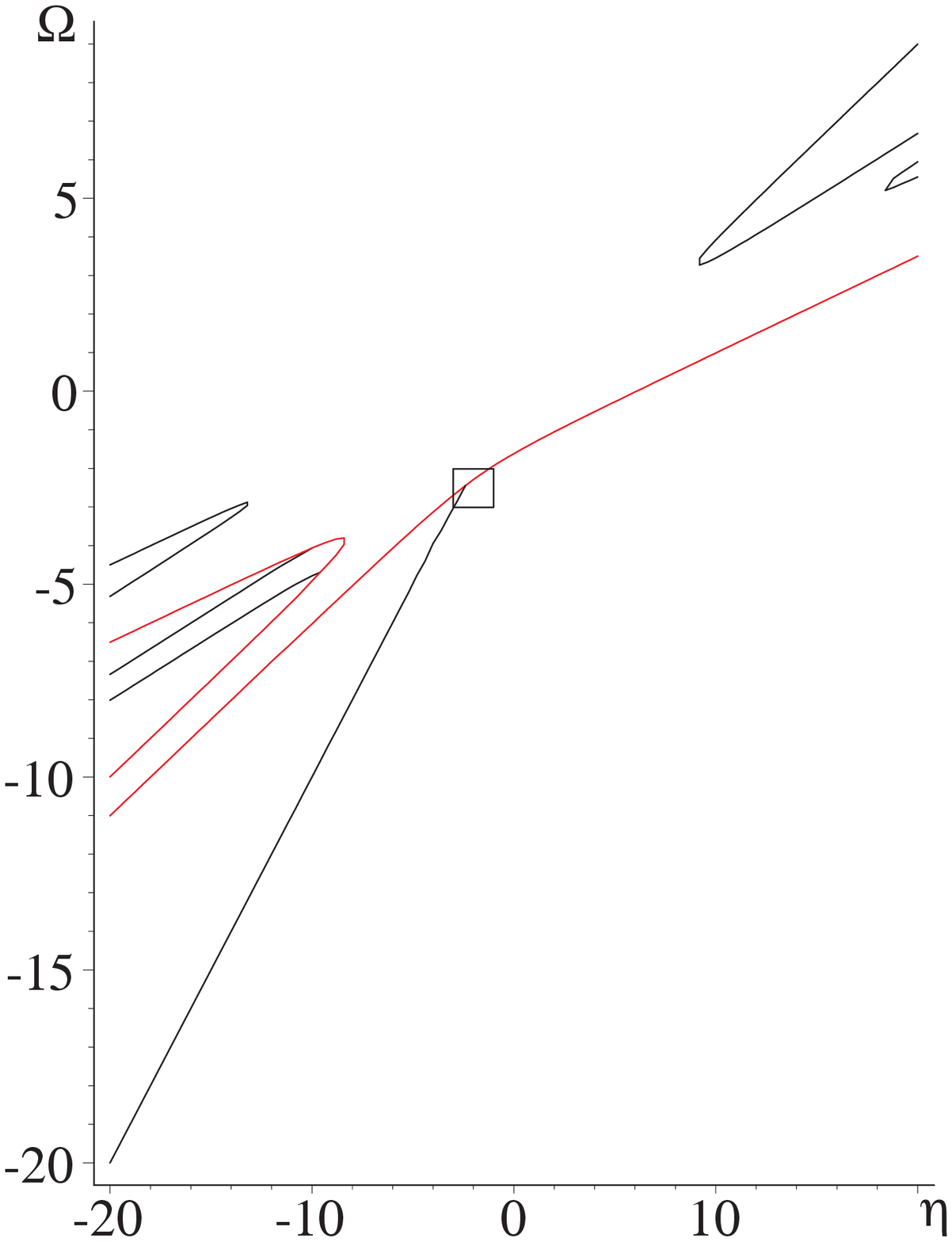}
\end{center}
\caption{\label {Fig6} Here we plot the energy $\Omega$ corresponding to symmetric (red lines) and asymmetrical (black lines) solutions versus $\eta$ in the case of cubic nonlinearity (i.e. $\sigma =1$). \ Here we choice $j=\ell =m=2$. \ The symmetric ground state bifurcates at $\eta =-2.31$ and new asymmetrical solutions occur. \ The bifurcation point, enclosed in the box, will be zoomed in Fig. \ref {Fig7}.}
\end{figure}

\subsection {Asymmetrical solutions} 
\begin{figure}[ht]
\begin{center}
\includegraphics[height=6cm,width=6cm]{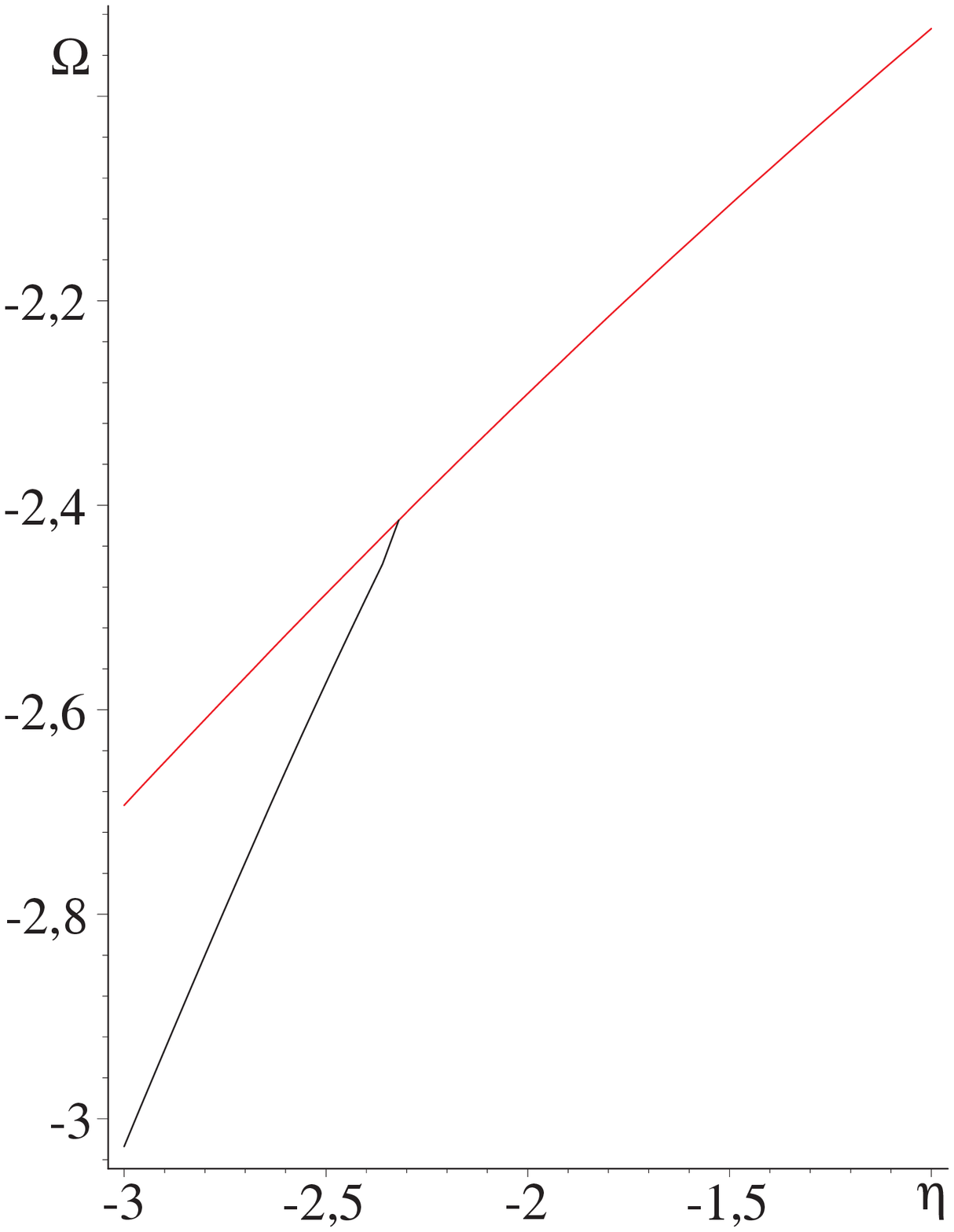}
\includegraphics[height=6cm,width=6cm]{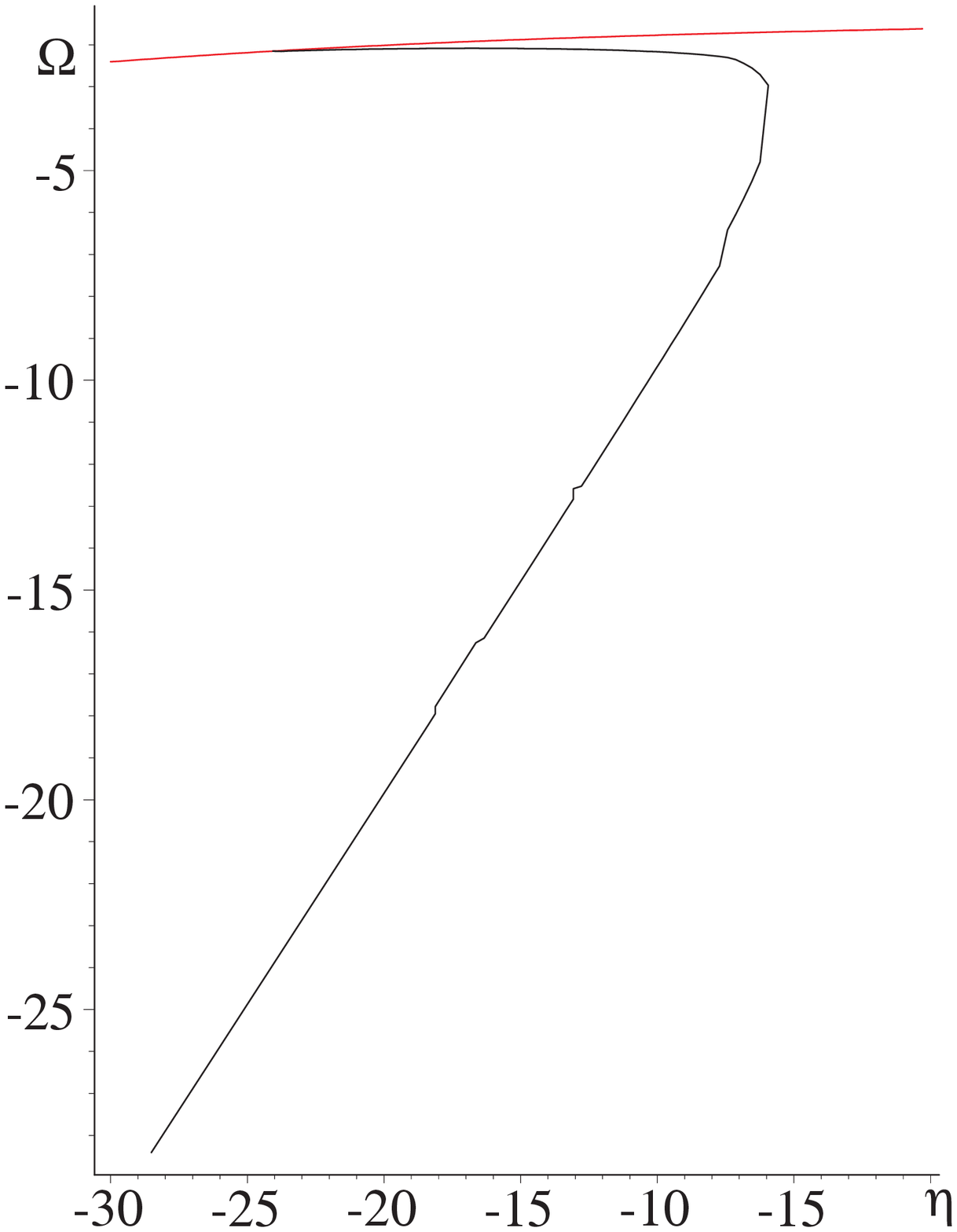}
\end{center}
\caption{\label {Fig7} In the left picture we zoom the bifurcation point in the cubic case, i.e. $\sigma =1$. \ For large value of $\sigma$ the kind of bifurcation may change; in the right picture we plot the bifurcation point for $\sigma=4$. \ Red lines denote the symmetric stationary solutions, black lines denote the branch of the asymmetrical stationary solutions.}
\end{figure}

Now, we look for the \emph {asymmetrical} solutions (\ref {Eq24FourTer}); actually, we have  a family of $8$ different systems of equations. \ In fact, we restrict our attention to the branches of asymmetrical solutions connected to the symmetric ground state. \ To this end we choose $j=\ell =m=2$. \ The numerical solutions of (\ref {Eq24FourTer}) in the cubic case (i.e. for $\sigma =1$) are plotted in Figure \ref {Fig6}; more precisely we plot the energy $\Omega$ as function of $\eta$. \ 
As appears in the piucture, branches of solution occurred when the effective nonlinearity parameter $\eta$ assumes critical values. \ Among these branches we restrict our attention to the branch with bifurcates from the symmetric stationary solution (we zoom the bifurcation in details in Figure \ref {Fig7} for different values of the nonlinearity power $\sigma$). \ 
As we can see a supercritical bifurcation of the symmetric stationary solution occurs at $\eta \approx -2.31$, the new branch behaves as $\Omega \approx \eta$ for large value of $|\eta |$, and the four almost-degenerate eigenfunctions are fully localized on one single well (as turns out in Table  \ref {Tab1}). \ 

\begin{table}[ht]
\begin{center}
\begin{tabular}{||c||c|c|c|c||} 
\hline
$\Omega $ & $q_1$ & $q_2$ & $q_3$ & $q_4 $ \\
\hline
\hline
-12.000 \ ($\mbox {e}_1$) & 0.5$\cdot 10^{-5}$ &0.007 & 0.986 & 0.007 \\ 
\hline
-12.000 \ ($\mbox {e}_2$) & 0.007 & 0.986 & 0.007 & 0.5$\cdot 10^{-5}$ \\ 
\hline
-11.999 \ ($\mbox {e}_3$) & 0.3$\cdot 10^{-6}$ & 0.5$\cdot 10^{-5}$ & 0.007 & 0.993 \\
\hline
-11.999 \ ($\mbox {e}_4$) & 0.993 & 0.007 & 0.5$\cdot 10^{-5}$ & 0.3$\cdot 10^{-6}$ \\
\hline
-7.021 \ (a)&  0.010 & 0.490 & 0.490 & 0.010 \\
\hline 
-7.009 & 0.2$\cdot 10^{-3}$ & 0.011 & 0.487 & 0.502 \\
\hline
-7.009 & 0.502 & 0.487 & 0.011 &0.2$\cdot 10^{-3}$ \\ 
\hline
-5.979 & 0.013 & 0.451 & 0.069 & 0.466 \\
\hline
-5.979 & 0.466 &  0.069 &  0.451 &  0.013 \\
\hline 
-5.952 \ (b)& 0.478 &  0.022 & 0.022 & 0.478 \\
\hline 
-5.376 &  0.013 & 0.364 &0.240 & 0.382 \\
\hline
-5.376 & 0.382 &0.240 &0.364 & 0.013 \\
\hline
-4.682 & 0.347 & 0.091 &0244 &0.317 \\ 
 \hline 
-4.682 & 0.317 & 0.244 &0.091 &0.347 \\ 
\hline 
-4.534 \ (c)&0.314 &0.186 & 0.186 &0.314 \\
\hline
\end {tabular}
\end {center}
\caption {Here we collect all the solutions corresponding to the value $\eta =-12$. \ Wavefunctions labeled with the letter (a), (b) and (c) coincides with the symmetrical ones already computed in (\ref {Eqpippo}); the other wavefunctions are asymmetrical wavefunctions. \ In particular the $4$ wavefunctions ($\mbox {e}_1$)-($\mbox {e}_1$) associated to the ground state $\Omega \sim - 12$ are fully localized on one single well} \label {Tab1}
\end {table}

We may remark that (see also Figure \ref {Fig7}, left panel) that the bifurcation is of the same supercritical kind as in the double well model (where bifurcation occurs at $\eta =-2$); furthermore, as in double well model, we observe (see also Figure \ref {Fig7}, right panel) a subcritical bifurcation of the symmetric stationary solution when the value of the nonlinearity power is bigger than the critical value  $\sigma_{threshold} = \frac {3+\sqrt{3}}{2}$ obtained by \cite {Sacchetti}. \ In adjoint to this spontaneous symmetry breaking effect of the ground state, which is the most relevant effect, other spontaneous symmetry breaking effect of the higher energy stationary states occur, and also new branches, associated to saddle points, of asymmetrical stationary states arise (see Figure \ref {Fig6} again).

\subsection {Ground state solution for large nonlinearity}

As discussed above, we have seen that for $|\eta |$ large enough (actually $\eta =-12$, as computed in Table \ref {Tab1}) the four almost-degenerate asymmetrical solutions, associated to the branch which bifurcates from the symmetric ground state, are localized on one single well. \ This result can be proved by means of a simple asymptotic argument as $\eta \to - \infty$. \ More precisely, let $j=m=\ell =2$ and let us set 
\bee
q_1 = 1 + \eta^{-2} s_1 \ \mbox { and } \ \Omega = \eta \left (1+ \eta^{-2} \Gamma \right )
\eee
where $s_1 =s_1 (\eta)$ and $\Gamma =\Gamma (\eta )$ will be discussed later. \ From equation (\ref {Eq24FourTer}-5) it follows that $q_k = \asy (\eta^{-2})$, $k=2,3,4$; more precisely, from equations (\ref {Eq24FourTer}-2), (\ref {Eq24FourTer}-3) and (\ref {Eq24FourTer}-4) immediately follows that 
\bee
q_2 = \eta^{-2} s_2 ,\ q_3 = \eta^{-4} s_3 \ \mbox { and } \ q_4 = \eta^{-6} s_4
\eee
where $s_k := s_k (\eta )$ are such that $s_4 \sim s_3 \sim s_2$, as $\eta \to - \infty$, and where $s_2 \sim 1 $, as $\eta \to - \infty$. \ Since equations (\ref {Eq24FourTer}-1) and (\ref {Eq24FourTer}-5) imply that
\bee
s_1 + s_2 \sim 0 \ \mbox { and } \ \sqrt {s2} + \sigma s_1 - \Gamma \sim 0 \, ,\ \mbox { as } \ \eta \to - \infty \, , 
\eee
hence
\bee
s_1 \sim -1 \ \mbox { and } \ \Gamma \sim 1-\sigma \, . 
\eee
This solution corresponds to the ground state, indeed it minimizes the Hamiltonian function (\ref {Eq18}) since $\sum_k q_k^{\sigma +1} \le \left [ \sum_k q_k \right ]^{\sigma +1}$ and $|\epsilon /\beta |$ is large enough (because we are considering the case $|\eta| \gg 1 $). 

Since a similar argument may apply when we choose, as starting point, $q_k = 1 + \eta^{-2} s_k$, for $k=2,3,4$, then we have proved the following result.

\begin {theorem}
There exists a value $\eta^\star >0$ such that the symmetrical stationary ground state bifurcates at $\eta= - \eta^\star$ and in the limit of large focusing nonlinearity, that is $\eta \to - \infty$, then and the four almost-degenerate asymmetrical solutions, which arise at the bifurcation point, are localized on one single well. 
\end {theorem}

\begin {remark}
We can extend this result to any number $N$ of wells; indeed the same asymptotic argument applies to the system (\ref {Eq22}) where we choose $\theta_k = \theta_{k+1}$ for any $k=1,2,\ldots , N-1$.
\end {remark}

\begin {remark}
By making use of the same arguments in \cite {RS} one may prove that the resulting almost-degenerate asymmetrical stationary solutions are orbitally stable; however we don't dwell here these details. 
\end {remark}

\section {Conclusion}

Semiclassical methods turn out to be a very powerfull tool in order to reduce a NLS to a finite-dimensional Hamiltonian systems. \ Indeed, by applying such techniques jointly with the Lyapunov-Schmidt reduction scheme, we are able to describe the ground state solutions as a superposition of vectors localized on single wells, with a rigorous estimate of the error \cite {RS}. \ In particular the Hamiltonian system (\ref {Eq17}) we obtain it is explicitely written, it can be reduced and it can be studied by means of standard numerical tools. 

In more details we consider the case with $N=4$ wells and we see that the spontaneous symmetry breaking effect, already observed in a double well model, similarly occurs; in particular we still observe supercritical bifurcation when the nonlinearity parameter is less that a \emph {threshold} value, for value bigger that such a threshold value a subcritical bifurcation occurs. 

A remarkable result is that in the case of large enough focusing nonlinearity then the symmetric ground state bifurcates and the new $N$ almost-degenerate stable  solutions are fully localized on one single well. \ This fact is very relevant from a physical point of view, indeed it is connected to, e.g. the explanation of the phase transition from superfluid to Mott-insulator state in the Bose-Hubbard model. \ Indeed, we can see in Fig. \ref {Fig6} a smooth transition from superfluidity (which corresponds to stationary solutions distributed on the whole lattice) to Mott insulator phase (which corresponds to stationary solutions localized on a single lattice cell without possibility to jump from one site to the others); the phase transition appears to be concentrated around to the values of $\eta_{bif}$ corresponding to the bifurcation point. \ in Table \ref {tabella1} we compute, for different values of the number $N$ of wells, the value of $\eta_{bif}$ for which we have a smooth transition from superfluidity to Mott insulator state, and we see that our results agree with the value of $\eta_{bif} \approx -1.8$ predicted by means of experimental calculation \cite {Stoferle} .

\begin{table}
\begin{center}
\begin{tabular}{|l||c|c|c|c|} 
\hline
$N $          &  2 & 4 & 6 &  8 \\ \hline 
$\eta_{bif}$      &  $-2.00$   & $-2.29$  &   $-2.37$ &  $-2.33$  \\ \hline
\end{tabular}
\caption{\small Table of values for $\eta$ at which the symmetric stationary solution for the Gross-Piatevskii equation, with lattice potential with $N$ wells, bifurcates.}
\label{tabella1}
\end{center}
\end {table}

\appendix

\section {Appendix}

Here we compute the eigenfunctions (\ref {Eq11Bis}) for the linear problem in the case of $N=2,3,4$ wells and where we assume, for argument's sake, the dimension $d=1$ and where the multiple well potential is given by a superposition of exactly equal and symmetric $N$ wells (i.e. $v(x)=v(-x)$). \ In sucha case the eigenvectors $\psi_j(x)$ are even and odd-parity functions.

In the case of \emph {two} wells, i.e. $N=2$, then the two eigenvalues are given by
\bee
\lambda_1 = \lambda_D - \beta \ \ \mbox { and } \ \ \lambda_2 = \lambda_D + \beta 
\eee
and the matrix $A=(\alpha_{j,k})$ has the form (in agreement with \cite {Sacchetti})
\bee
A = \frac {\sqrt{2}}{2}
\left ( 
\begin {array}{cc}
1 & 1 \\
1& -1
\end {array}
\right ) .
\eee
In the case of \emph {three} wells, i.e. $N=3$, then 
\bee
\lambda_1 = \lambda_D - \sqrt {2} \beta \, ,\ \lambda_2 = \lambda_D \ \ \mbox { and } \ \ \lambda_3 = \lambda_D + \sqrt {2} \beta 
\eee
and (in agreement with \cite {Kapitula})
\bee
A = \frac 12 \left ( 
\begin {array}{ccc}
1 & \sqrt{2} & 1 \\
\sqrt{2} & 0 & -\sqrt{2} \\ 
1 & - \sqrt{2} & 1
\end {array}
\right ) .
\eee
In the case of \emph {four} wells, i.e. $N=4$, then 
\bee
\lambda_1 &=& \lambda_D - 2 \beta \cos \left ( \frac {\pi}{5} \right ) \, ,\ \lambda_2 = \lambda_D - 2 \beta \cos \left ( \frac {2\pi}{5} \right ) \\ 
\lambda_3 &=& \lambda_D + 2 \beta \cos \left ( \frac {2\pi}{5} \right ) \, ,\ \lambda_4 = \lambda_D + 2 \beta \cos \left ( \frac {\pi}{5} \right ) 
\eee
and 
\bee
A = \frac {\sqrt {10}}{5} 
\left ( 
\begin {array}{cccc}
\sin (\pi/5) & \sin (2\pi/5) & \sin (2\pi/5) & \sin (\pi/5) \\
\sin (2\pi/5) & \sin (\pi/5) & -\sin (\pi/5) & -\sin (2\pi/5) \\
\sin (2\pi/5) & -\sin (\pi/5) & -\sin (\pi/5) & \sin (2\pi/5) \\
\sin (\pi/5) & -\sin (2\pi/5) & \sin (2\pi/5) & -\sin (\pi/5)
\end {array}
\right ) .
\eee

For instance, see Figures 1, 2 and 3 for the one-dimensional $N$-wells problem with, respectively, $N=2$, $N=3$ and $N=4$; where the wavefunctions $\psi_D(x)$ and $\psi_j (x)$ can be chosen to be real-valued functions.

\begin{figure}
\begin{center}
\includegraphics[height=5cm,width=5cm]{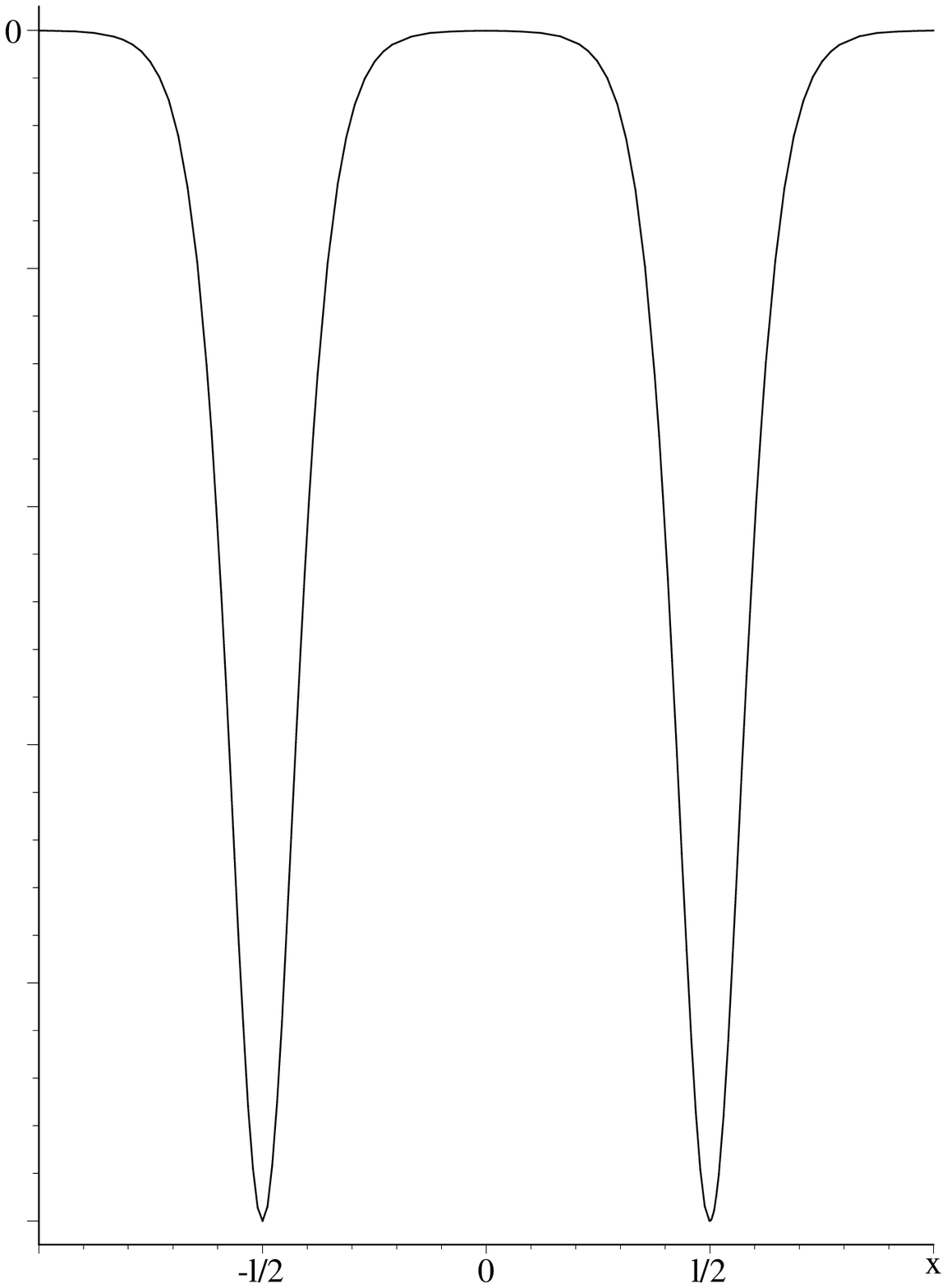}
\includegraphics[height=5cm,width=5cm]{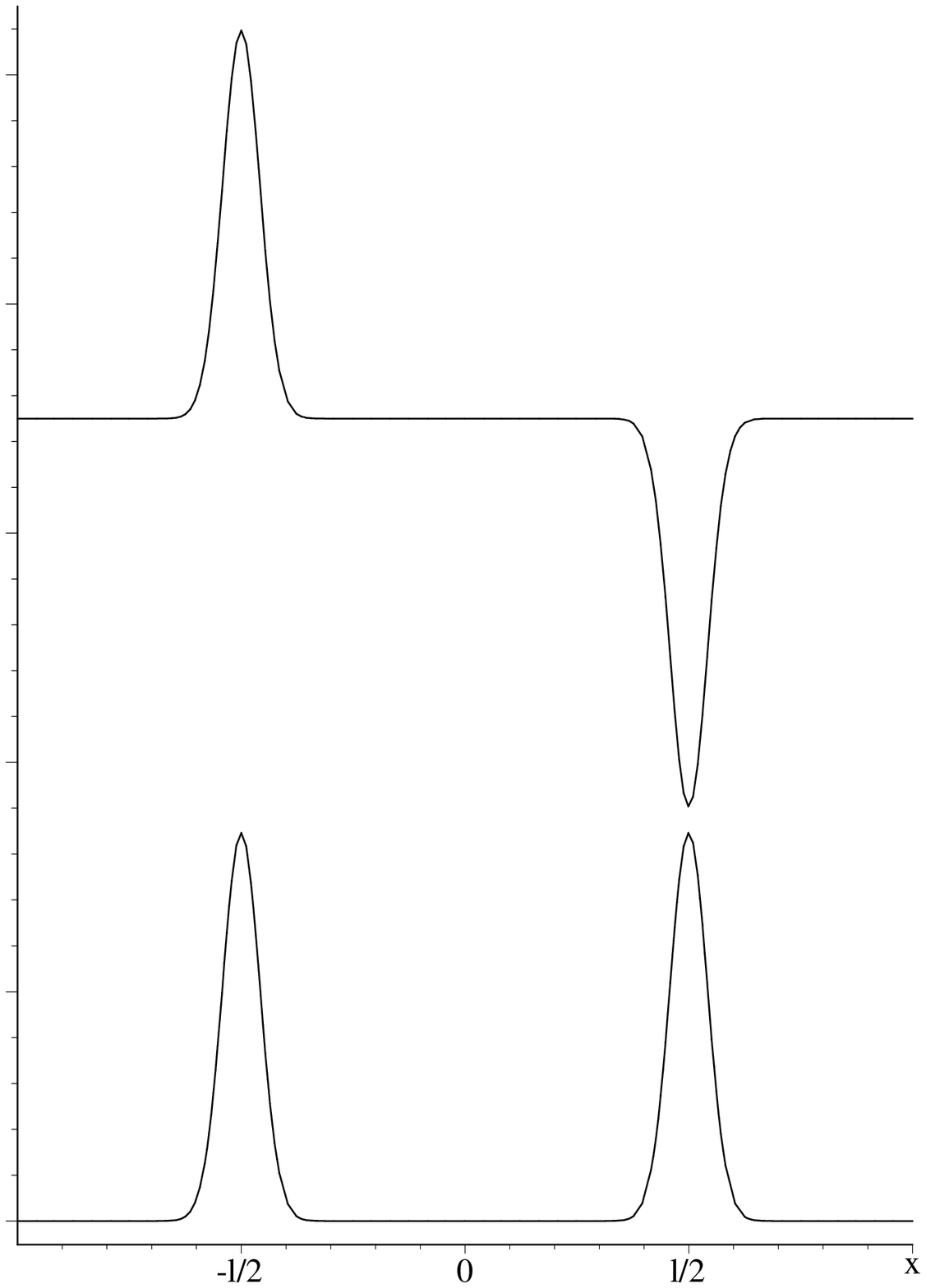}
\end{center}
\caption{\label {Fig1} Double-well potential (left) and the two symmetric and antisymmetrical eigenvectors (right).}
\end{figure}

\begin{figure}
\begin{center}
\includegraphics[height=5cm,width=5cm]{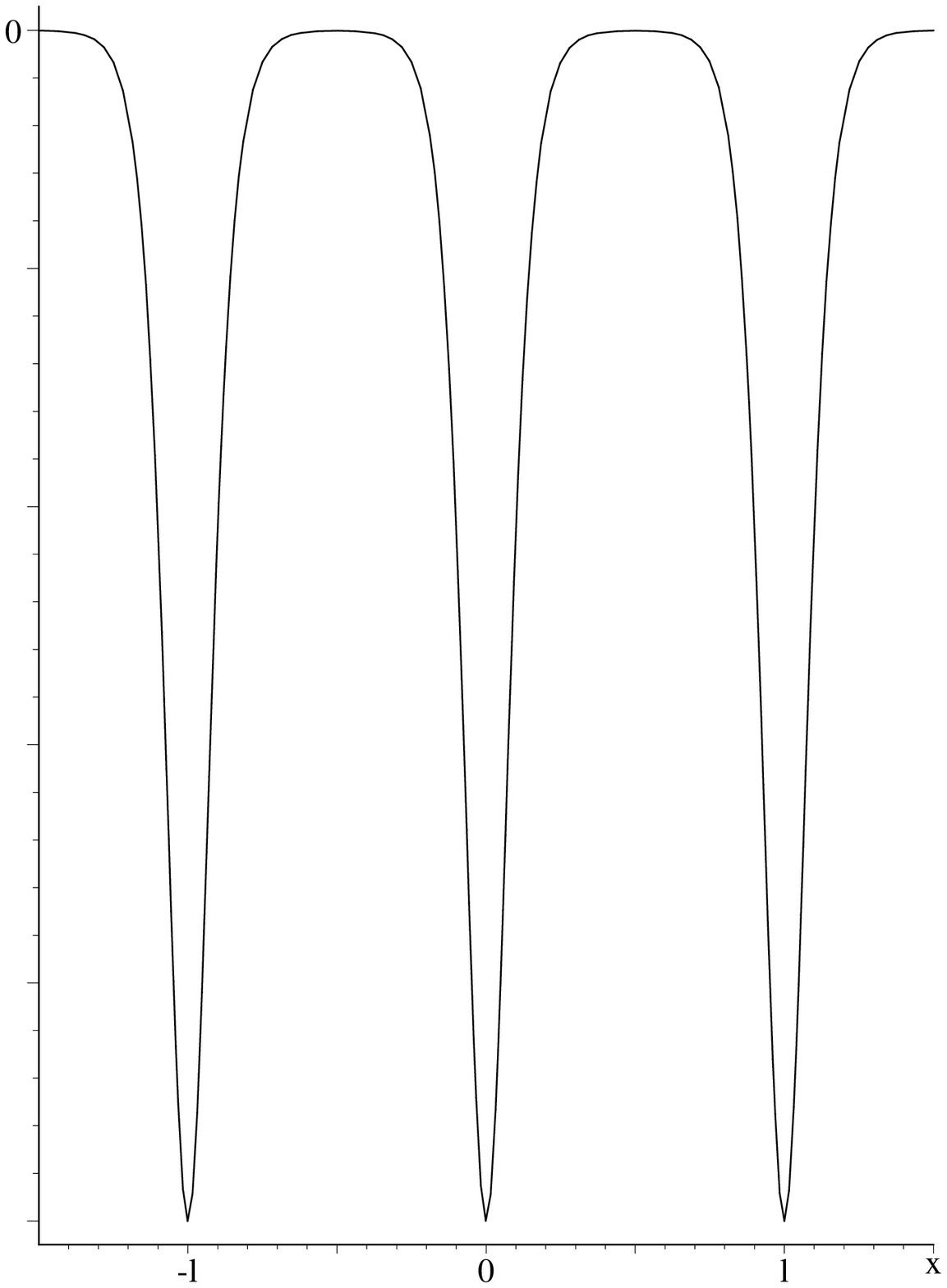}
\includegraphics[height=5cm,width=5cm]{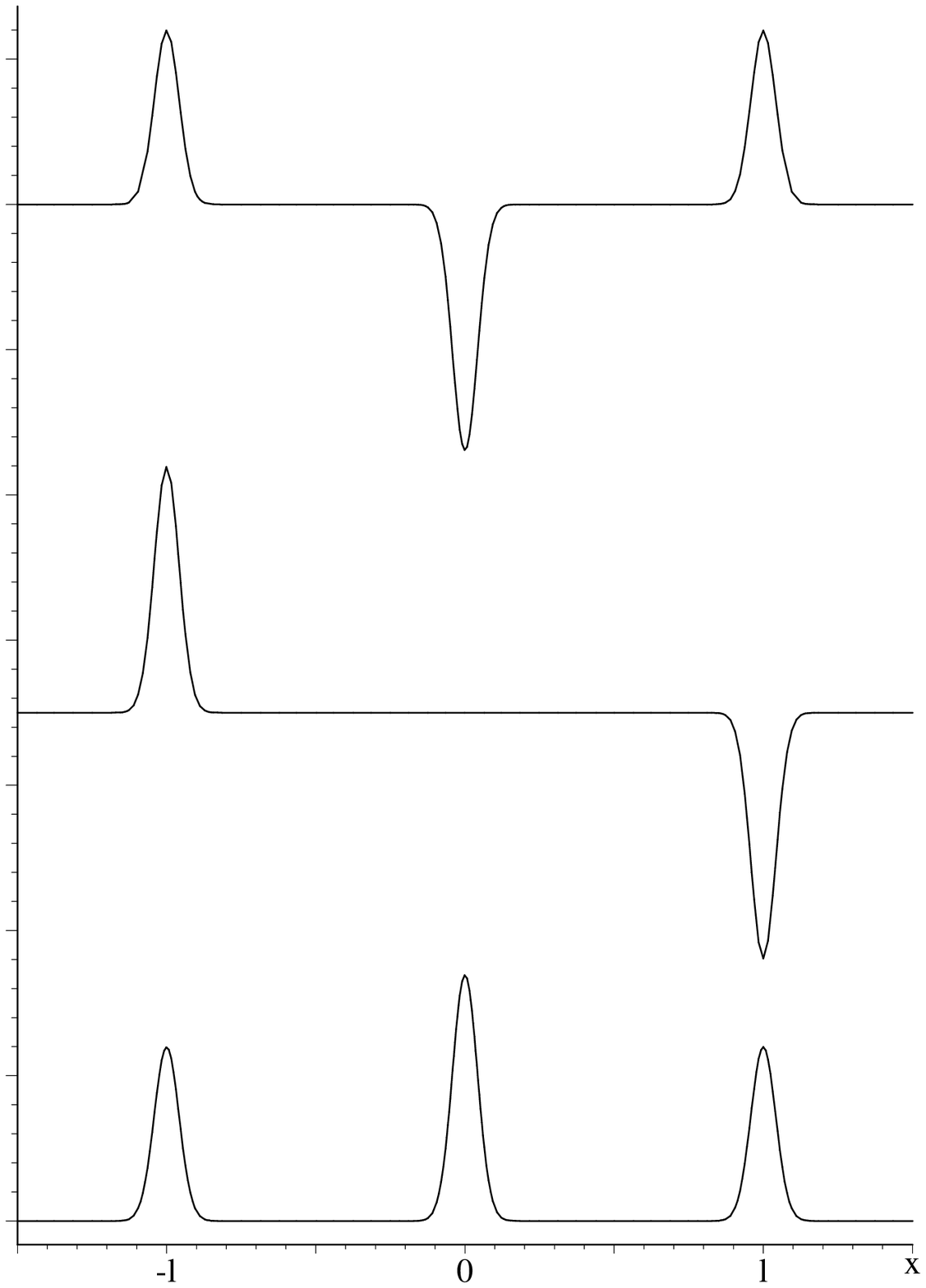}
\end{center}
\caption{\label {Fig2} Three-well potential (left) and the three symmetric and antisymmetrical eigenvectors (right).}
\end{figure}

\begin{figure}
\begin{center}
\includegraphics[height=5cm,width=5cm]{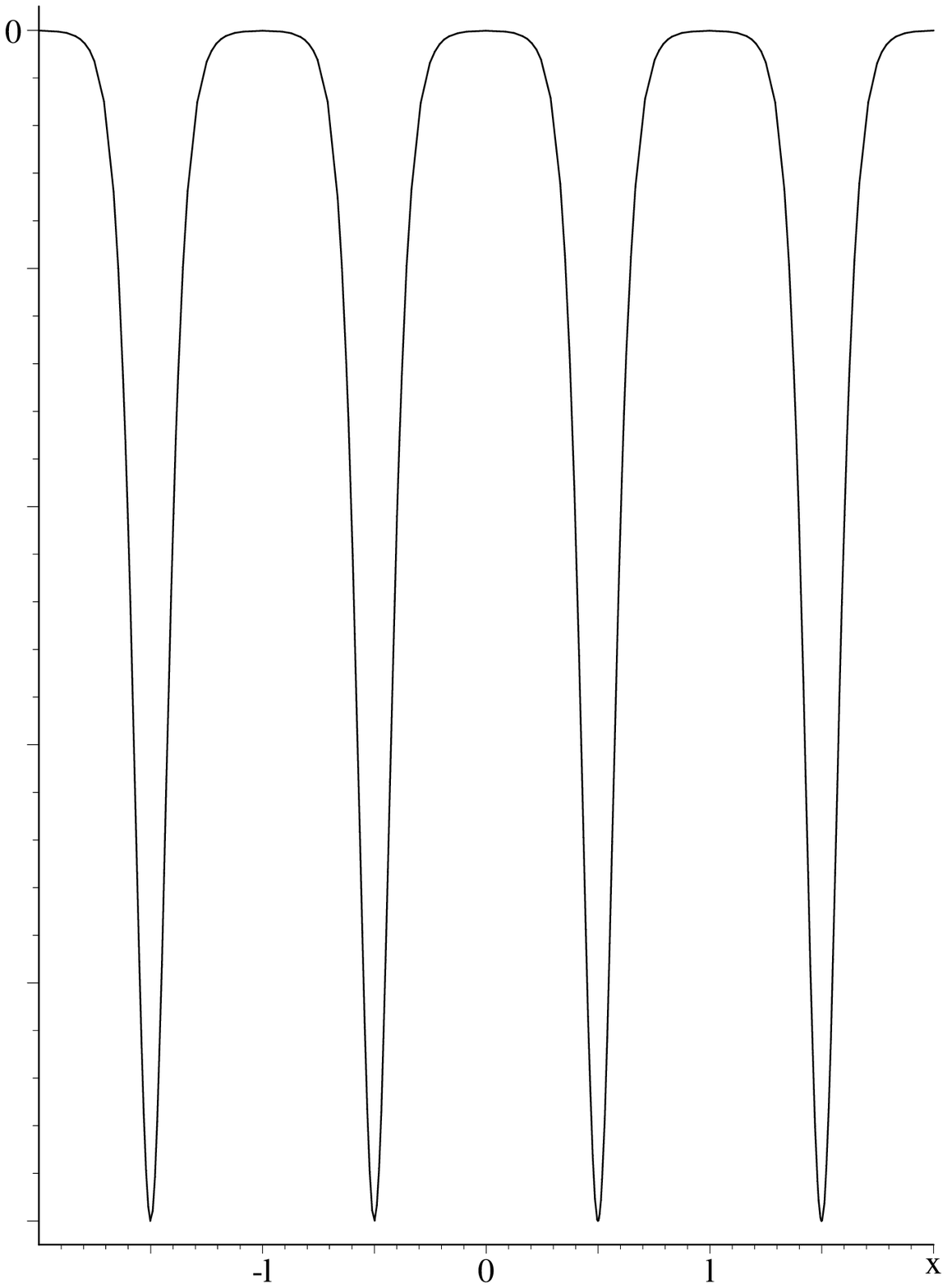}
\includegraphics[height=5cm,width=5cm]{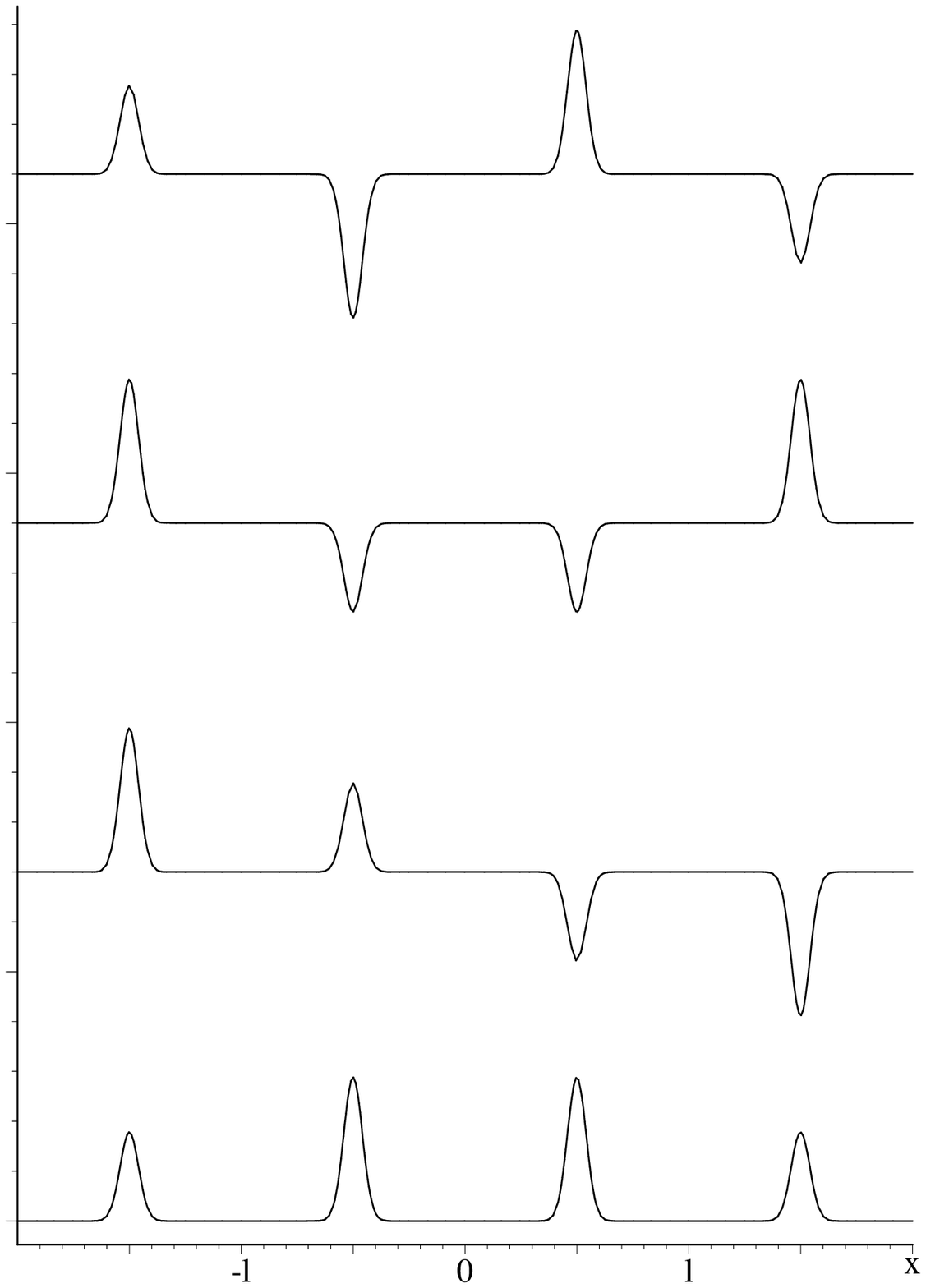}
\end{center}
\caption{\label {Fig3} Four-well potential (left) and the four symmetric and antisymmetrical eigenvectors (right).}
\end{figure}


\begin{thebibliography}{99}

\bibitem {A} Anderson, M.H., Ensher, J.R., Matthews, M.R., Wieman, C.E. and Cornell, E.A. \ {Observation of Bose-Einstein Condensation in a Dilute Atomic Vapor}. {\it Science} {\bf 269}, 198-201 (1995).

\bibitem {BS} D.Bambusi, and A.Sacchetti, {\it Exponential times in the one-dimensional Gross-Pitaevskii equation with multiple well potential}, Comm. Math. Phys. {\bf 275}, 1-36 (2007).

\bibitem {Bl} Bloch I. \ Ultracold quantum gases in optical lattices. {\it Nature Physics} {\bf 1}, 23-30 (2005).

\bibitem {BSH} Bradley, C.C., Sackett, C.A. and Hulet, R.G. \ {Bose-Einstein condensation of lithium: Observation of limited condensate number}. {\it Phys. Rev. Lett.} {\bf 78}, 985-989  (1997).

\bibitem {RS} R.Fukuizumi, and A.Sacchetti, {\it Bifurcation and stability for Nonlinear Schr\"odinger equations with double well potential in the semiclassical limit}, preprint (2011)

\bibitem {GW} Z.Gang, and M.I.Weinstein, {\it Equipartition of mass in nonlinear Schr\"dinger/Gross-Pitaevskii equations}, Appl. Math. Res. Express {\bf 2011}, 123-181 (2011).

\bibitem {Grecchi} V.Grecchi, and A.Martinez, {\it Non-linear Stark effect and molecular localization}, Comm. Math. Phys. {\bf 166}, 533-548 (1995).

\bibitem {GMS} V.Grecchi, A.Martinez, and A.Sacchetti, {\it Destruction of the beating effect for a non-linear Schr\"odinger equation}, Comm. Math. Phys. {\bf 227}, 191-209 (2002).

\bibitem {HMEWC} Hall, D.S., Mattthews, M.R., Ensher, J.R., Wieman, C.E. and Cornell, E.A. \ {Dynamics of component separation in a binary mixture of Bose-Einstein condensates}. {\it Phys. Rev. Lett.} {\bf 81}, 1539-1542 (1998).

\bibitem{H} B.Helffer, {\it Semi-classical Analysis for the Schr\"odinger operator and applications}, Lecture Note in Mathematics, 1336, Springer-Verlag (1980).

\bibitem {JaWe} R.K.Jackson, and M.I.Weinstein, {\it Geometric Analysis of Bifurcation and Symmetry Breaking in a Gross-Pitaevskii Equation}, J. Stat. Phys. {\bf 116}, 881-905 (2004).

\bibitem {Joannopoulos} J.D.Joannopoulus, S.G.Johnson, J.N.Winn, and R.D.Meade, {\it Photonic Crystals: molding the flow of light}, (Princeton Univ. Press: 2008)

\bibitem {Jona} G.Jona-Lasinio, C.Presilla, and C.To\-ni\-nel\-li, {\it Interaction induced localization in a gas of py\-ra\-mi\-dal molecules}, Phys.Rev.Lett. {\bf 88}, 123001  (2002).

\bibitem {Kapitula} T.Kapitula, P.G.Kevrekidis, and Z.Chen, {\it Three is a crowd: solitary waves in photorefractive media with three potential wells}, SIAM J. Appl. Dyn. Syst. {\bf 5}, 598-633 (2006).

\bibitem {Peli} E.W. Kirr, P.G. Kevrekidis, and D.E. Pelinovsky, {\it Symmetry-breaking bifurcation in the nonlinear Schrodinger equation with symmetric potentials}, Communications in Mathematical Physics (2011).

\bibitem {KKSW} E.W.Kirr, P.G.Kevrekidis, E.Shlizerman, and M.I.Weinstein, {\it Symmetry-breaking bifurcation in nonlinear Schr\"odinger/Gross-Pitaevskii equations}, SIAM J. Math. Anal. {\bf   40}, 566-604 (2008).

\bibitem {MW} J.Marzuola, and M.I.Weinstein, {\it Long time dynamics near the symmetry breaking bifurcation for nonlinear Schr\"odinger / Gross-Pitaevskii equations}, Discrete and Continuous Dynamical Systems B {\bf 28}, 1505-1547 (2010).

\bibitem {M} C.D.Meyer, {\it Matrix analysis and applied linear algebra}, SIAM (2004).

\bibitem {Mihalace} D.Mihalace, M.Bertolotti, and C.Sibilia, {\it Nonlinear wave propagation in planar structures}, Prog. Opt. {\bf 27}, 229 (1989)).

\bibitem {Pitaevskii} L.Pitaevskii, and S.Stringari, {\it Bose-Einstein condensation}, (Claredon Press: Oxford 2003). 

\bibitem {S} A.Sacchetti, {\it Nonlinear double well Schr\"odinger equations in the semiclassical limit}, J. Stat. Phys. {\bf 119}, 1347-1382 (2005).

\bibitem {Sacchetti} A.Sacchetti, {\it Universal critical power for nonlinear Schr\"odinger equations with a symmetric double well potential}, Phys. Rev. Lett. {\bf 103}, 194101 (2009).

\bibitem {Stoferle} St\"oferle, T., Moritz, H., Schori, C., K\"ohl, M. and Esslinger T. \ Transition from a Strongly Interacting 1D Superfluid to a Mott Insulator. {\it Phys. Rev. Lett.} {\bf 92} (2004), 130403:1-4.

\bibitem {Wang} C.Wang, G.Theocharis, P.G.Kevrekidis, N.Whitaker, K.J.H.Law, D.J.Frantzeskakis, and B.A.Malomed, {\it Two-dimensional paradigm for symmetry breaking: the nonlinear Schr\"odinger equation with a four-well potential}, Phys. Rev. E {\bf 80}, 046611 (2009).

\end{thebibliography}
\end {document}